\documentclass[aps,prd,showpacs,superscriptaddress,nofootinbib,10pt]{revtex4-2}

\usepackage{acronym}
\usepackage{amsfonts}
\usepackage{amsmath}

\usepackage{amssymb}
\usepackage{bm}
\usepackage{cases}
\usepackage{color}
\usepackage{comment}
\usepackage[dvipsnames]{xcolor}
\usepackage{enumerate}
\usepackage{epsfig}
\usepackage{etoolbox}
\usepackage{fancyref}
\usepackage{fancyhdr}
\usepackage[T1]{fontenc} 
\usepackage{graphicx}
\usepackage{hyperref}
\usepackage{indentfirst}
\usepackage{latexsym}
\usepackage{mathrsfs}
\usepackage{mciteplus}
\usepackage{multirow}
\usepackage{rotating}
\usepackage{scrextend}
\usepackage{soul}
\usepackage{subfigure}
\usepackage{verbatim}
\usepackage{amsthm} 
\usepackage{tabularx}
\usepackage{enumitem}    
\usepackage{array}

\begin{document}

\title[Master variables and Darboux symmetry for axial perturbations of the exterior and interior of black hole spacetimes
]{Master variables and Darboux symmetry for axial perturbations of the exterior and interior of black hole spacetimes}

\author{Michele Lenzi}
\email{michele.lenzi@ulb.be}
\affiliation{Physique Théorique et Mathématique, Université Libre de Bruxelles (ULB), Campus Plaine, Building NO, CP 231, Boulevard du Triomphe B-1050, Bruxelles, Belgique}

\author{Guillermo A. Mena Marug\'{a}n}
\email{mena@iem.cfmac.csic.es}

\author{Andr\'{e}s M\'{ı}nguez-S\'{a}nchez}
\email{andres.minguez@iem.cfmac.csic.es}
\altaffiliation{Affiliated to the PhD Program, Departamento de F\'isica Te\'orica, Universidad Complutense de Madrid, 28040 Madrid, Spain}
\affiliation{Instituto de Estructura de la Materia, IEM-CSIC, Serrano 121, 28006 Madrid, Spain}

\author{Carlos F. Sopuerta}
\email{carlos.f.sopuerta@csic.es}
\affiliation{Institut de Ci\`encies de l'Espai (ICE, CSIC), Campus UAB, Carrer de Can Magrans s/n, 08193 Cerdanyola del Vall\`es, Spain}
\affiliation{Institut d'Estudis Espacials de Catalunya (IEEC), Edifici Nexus, Carrer del Gran Capit\`a 2-4, despatx 201, 08034 Barcelona, Spain}


%
%
\begin{abstract}
Recent efforts have shown that Kantowski-Sachs spacetime provides a useful framework for analyzing perturbations inside a Schwarzschild black hole (BH). In these studies, the adoption of a Hamiltonian formulation offers an insightful perspective. The aim of this work is twofold. First, we revisit and elaborate the results obtained so far in Kantowski-Sachs spacetime, with the focus placed on axial perturbations. In particular, by exploiting the relation between this spacetime and the interior of a nonrotating BH, we consider the extension of those results to the exterior geometry of the BH. In this way, we clarify the relation between the axial perturbative gauge invariants emerging from the canonical analysis and the already well-established axial BH invariants, often referred to as master functions. We do so by providing a unified picture of the Hamiltonian formalism, which does not distinguish, formally, between exterior and interior geometries. The second objective is to explore the role of Darboux transformations, which were found as hidden symmetries in the context of BH perturbations, and their appearance in the Hamiltonian setting. Within this framework, the Hamiltonian formulation provides a clear geometric interpretation and characterization of Darboux transformations within the axial sector, viewing them as the set of canonical transformations between Hamiltonians for axial master functions.
\end{abstract}

\maketitle

%
%

\section{Introduction}

Perturbation theory is a fundamental tool in physics with an exceptionally wide range of applications. General relativity (GR) is one of the many fields where it plays a central role. Its use spans several areas of astrophysics and cosmology, including studies of the origin and evolution of cosmic microwave background anisotropies~\cite{Bardeen:1980PhRvD..22.1882B,Mukhanov:1990me,Planck:2018nkj}, the structure of relativistic stars and black holes (BH), and the analysis and propagation of gravitational waves (GWs)~\cite{Chandrasekhar:1992bo}. These topics remain central in modern physics, and their theoretical foundations rely heavily on perturbative methods. In the context of gravitational perturbations over BH backgrounds, perturbation theory provides the natural framework to study several physical properties of the spacetime itself. These range from the study of scattering of waves and particles from BHs~\cite{Futterman:1988ni}, providing the BH graybody factors, central for the description of Hawking radiation~\cite{Hawking:1975vcx}, to the definition of the so called BH quasinormal modes~\cite{Nollert:1999re,Kokkotas:1999bd,Berti:2009kk,Konoplya:2011qq} and tidal deformations~\cite{Damour:2009vw,Binnington:2009bb}. In the GW era, these quantities are important smoking guns for possible deviations from the general relativistic BH paradigm, some of which may be testable by future GW detectors such as LISA~\cite{LISA:2017pwj, LISA:2022kgy} and the Einstein Telescope~\cite{ET:2025xjr}.

The first pioneering works in this context studied perturbations of an uncharged nonrotating BH. As is customary in systems with spherical symmetry, one can expand the perturbations into scalar, vector, and tensor spherical harmonics of odd and even parity. This decomposition isolates the angular dependence, and the equations for the perturbations then decouple into distinct harmonic modes. For each harmonic multipole and parity, one can identify particular combinations of the metric perturbations, called master functions, which satisfy a master equation with a potential. The master functions are first order gauge invariants constructed from a suitable linear combination of the metric perturbations and their derivatives. The master equation is a $1+1$-dimensional wave-type equation (in a two dimensional Lorentzian geometry) that governs the dynamics of the physically relevant perturbations. This was first shown in Schwarzschild spacetime by Regge and Wheeler~\cite{Regge:1957td} and Cunningham, Price, and Moncrief (CPM)~\cite{Cunningham:1978cp} who found different, but related, odd/axial master functions and corresponding master equations. The even/polar case was first decoupled by Zerilli and Moncrief (ZM)~\cite{Zerilli:1970se,Moncrief:1974vm}. The extension to any spherically symmetric spacetime has been shown by Gerlach and Sengupta (GS)~\cite{Gerlach:1979rw,Gerlach:1980tx}. It has recently been shown in Refs.~\cite{Lenzi:2021wpc, Lenzi:2024tgk} that there are actually infinite physically equivalent ways to decouple Einstein equations for the perturbations into master equations on any spherically symmetric spacetime (with cosmological constant). This has been understood by the presence of a hidden symmetry in BH perturbation theory called Darboux covariance~\cite{Lenzi:2021njy}.

In this work, we consider a Kantowski-Sachs spacetime, a geometry used to describe the interior of a Schwarzschild BH, and study the first order axial perturbations in a fully Hamiltonian framework. The Hamiltonian formulation for perturbations of vacuum spherically symmetric spacetimes was pioneered by Moncrief~\cite{Moncrief:1974vm} and later extended to dynamical backgrounds in Refs.~\cite{BrizuelaPhD, Brizuela:2008sk}. The perturbations of Kantowski-Sachs backgrounds were studied with Hamiltonian methods in Refs.~\cite{MenaMarugan:2024qnj,MenaMarugan:2025anx}. In the Hamiltonian formulation of GR, one adopts a 3+1 decomposition of the spacetime, introducing a foliation of the four-dimensional manifold into an evolution direction and three-dimensional slices, according to the proposals of Arnowitt, Deser, and Misner (ADM)~\cite{Arnowitt:1962hi}. From a Hamiltonian perspective, GR is degenerate because of the covariant nature of the theory, in the sense that there is no (bulk) Hamiltonian functional that strictly generates something completely equivalent to a time evolution. Instead, the system is entirely constrained, and the constraints generate the true dynamics. Rather than working directly with the full spacetime metric, the Hamiltonian formalism decomposes the gravitational degrees of freedom into the metric induced on the three-dimensional sections of the foliation, the lapse function, and the shift vector (which together reconstruct the full spacetime metric), along with their corresponding conjugate momenta. Practically, only the spatial metric, its momentum, the lapse and shift, and the constraints that generate spacetime diffeomorphisms, play an essential role in the canonical formulation. The Hamiltonian approach to perturbation theory modifies this structure only partly. To leading order, the perturbed phase space is built from the linear contributions of the canonical variables. In fact, this provides a presymplectic structure that is already quadratic in the perturbations. In addition, the contribution of the smeared constraints to the total gravitational action must be expanded also to second order (the same order as the contribution of the perturbative Legendre terms that accounts for the aforementioned presymplectic structure). These second order pieces are special: they contain no genuine second order perturbations, but arise solely from products of first order perturbations, including those of the lapse and shift. The resulting smeared constraints govern the first order dynamics.

Here, we use the Hamiltonian procedure developed in the context of Kantowski-Sachs spacetimes in Refs.~\cite{MenaMarugan:2024qnj,MenaMarugan:2025anx} with a twofold perspective. On the one side we aim at revisiting the study of perturbations in the (analytically continued) interior of a Schwarzschild BH and discuss the steps that allow us to connect to the well-known gauge invariant formulation usually adopted within the context of perturbation theory of spherically symmetric spacetimes. In this study, we focus our attention on axial perturbations. On the other hand, at least for axial perturbations, our approach allows us to provide an ``off-shell'' Hamiltonian perspective on the Darboux hidden symmetry of Refs.~\cite{Lenzi:2021wpc,Lenzi:2021njy,Lenzi:2024tgk}, which is understood as a canonical symmetry. Starting from an ADM Hamiltonian for the perturbations over our curved backgrounds, we can identify three requirements, necessary to obtain the desired results starting from quite general motivations: i) the gauge invariance of the dynamical variables; ii) the diagonalization of the Hamiltonian; iii) constant terms in front of squared momenta, so that these momenta are just the dynamical derivatives of the configuration variables. These three steps will comply with both previously mentioned objectives. It is important to stress that the splitting into these three different steps is meant to highlight the central features and motivations, and to show how one can systematically obtain the desired Hamiltonians, without having to make educated guesses driven by the \textit{a priori} knowledge of the final result. However, since every step is represented by a canonical transformation, one can compose them and jump directly to the final result, once the procedure is understood. In particular, one should pay attention to avoid the inclusion of possible nontrivial and nonlocal physics in this artificial division into steps, as we will argue in this development. In following this procedure, we will keep the formalism general enough so that it describes at once the Schwarzschild interior and exterior geometries. Indeed, with a complex transformation of the tetrad-connection variables chosen to describe the background, the formalism developed for Kantowski-Sachs spacetime can be extended to the exterior region as well. Since this transformation acts on the background phase space rather than on coordinates, it should not be taken as a conventional Wick rotation. The only difference between interior and exterior regions is related to the evolution parameter, which is a time coordinate in the interior but a radial one in the exterior. The use of radial Hamiltonians in spherically symmetric systems has already appeared in a number of contexts recently~\cite{BenAchour:2022uqo,BenAchour:2023dgj,Perez:2023jrq,Livine:2025soz}.

The structure of the article is as follows. In Sec.~\ref{S:Background} we introduce the background, namely a Kantowski-Sachs spacetime in triad and connection variables, and show how this can provide a unified description of the Schwarzschild exterior and its interior analytic continuation within the same mathematical formulation. 

In Sec.~\ref{S:Perturbative-action} we use the ADM formalism, adapted to spherical symmetry, to obtain the second order action describing the dynamics of first order perturbations in a Hamiltonian formalism, thus isolating dynamical variables, gauge variables, and constraints. We concentrate our discussion on axial perturbations for concreteness and simplicity. Starting from this, in Subsec.~\ref{Ss:perturbative-GI} we review how to obtain the action for axial gauge invariants without focusing on further requirements on the form of the Hamiltonian, which is in general nondiagonal and contains background dependent factors in all its quadratic perturbative terms. 

In Sec.~\ref{S:darboux-symmetry} we show how Darboux transformations (DTs) emerge in the Hamiltonian framework for axial perturbations. We first introduce, in Subsec.~\ref{Ss:diagonalizing-canonical}, the formal mathematical tools that we use to manipulate the Hamiltonians obtained in the previous section into diagonal Hamiltonians with no (nontrivial) background dependence in front of the squared momenta. The diagonal requirement (useful for a quantization of the perturbations) removes ambiguities in the definition of the canonical momentum variables, while the other requirement guarantees that these momenta are the dynamical derivatives of the configuration variables. While providing a useful scheme, these transformations are still formal because we consider an extremely general framework that may contain undesired nonstandard physics owing to nonlocality. These potential problems will be avoided when considering the actual physical scenario of perturbations in Kantowski-Sachs spacetimes. In Subsec.~\ref{Ss:Darboux-covariance} we define what we mean by DTs in an on-shell perspective, i.e. at the level of the equations of motion. In Subsec.~\ref{Ss:darboux-canonical} we show that the Hamiltonians obtained by the diagonalization procedure in Subsec.~\ref{Ss:diagonalizing-canonical} are actually related among each other by the so-called generalized DTs among axial perturbations. These contain the genuine axial DTs as the physically relevant subclass, in the sense that they avoid nonconventional nonlocality. In this way, we provide the correct off-shell canonical representation of the on-shell DTs.

In Sec.~\ref{S:KS-Darboux} we study the relation of the Hamiltonians for the axial perturbations in Kantowski-Sachs spacetimes to Hamiltonians for the typical axial master functions considered in BH perturbation theory. We apply in Subsec.~\ref{Ss:KS-master} the formal constraints introduced in Subsec.~\ref{Ss:diagonalizing-canonical} to show that the diagonalization procedure, subject to the physical requirement of frequency independence of the potential, selects a family of Darboux related Hamiltonians, in the sense described in Subsec.~\ref{Ss:darboux-canonical}. A special element of this class is the Hamiltonian for the CPM master function with the Regge-Wheeler potential, as shown explicitly in Subsec.~\ref{Ss:RW-equation}. Finally, we show the relation with the GS axial master variable in Sec.~\ref{Ss:GS-master}, which turns out to be obtained through a scaling in the configuration variables, combined with a suitable transformation in the canonical momenta that keeps the diagonal character and the background independent nature of the coefficients of the squared momenta. The above canonical transformation implies  a redefinition of the evolution parameter between the CPM and GS descriptions which produces a Schwarzian derivative term in the potential.

Finally, Sec.~\ref{conclusion} contains the conclusion and some further comments. Three appendices are added with long formulas of the Hamiltonian construction and some technical details about Laplace-Beltrami operators. 

In our discussion, we adopt geometric units such that the Newton constant and the speed of light are equal to the unit. In addition, Greek letters from the middle of the alphabet are used for spacetime indices, the Greek letters $\chi$ and $\zeta$ denote coordinates and indices in the Lorentzian two-dimensional manifold (which describes the set of spherical orbits), capital Latin letters (as well as the letters $\theta$ and $\phi$) are used for indices in the 2-sphere, and lowercase Latin letters  $i,j,k,...$ are the three (internal) spatial indices, unless explicitly stated otherwise.

\section{Background geometries}
\label{S:Background}

In this Section we introduce the background Hamiltonian formulation of the homogeneous but anisotropic Kantowski-Sachs spacetime that represents the zero order description in the perturbative framework. One of the main advantages of working within a Hamiltonian framework is that this allows us to keep the background mostly general, only requiring its spherical symmetry together with the existence of a Killing vector that is hypersurface orthogonal. Note that, in principle, one could even consider a phenomenological background and study the perturbations over it, assuming their dynamics are given by the linearized Einstein theory. On the other hand, we will sometimes fix the background equal to the analytic extension within the horizon of the Schwarzschild solution to GR, for the purpose of making clear comparison with the known results from BH perturbation theory.

\subsection{Background in triad-connection formulation}

The Hamiltonian description involves: 1) the identification of the canonical pairs (or Hamiltonian variables), which characterize the symplectic structure (also expressed through the Poisson bracket) and thus define the phase space; 2) the identification of the constraints that restrict the canonical pairs; and 3) if nonvanishing, the introduction of the physical Hamiltonian itself. The latter two elements are responsible for generating the dynamical evolution of the system. In the case of GR, the physical Hamiltonian can at most be given by surface terms, which by no means affect the dynamical equations on the bulk.

In GR, the background canonical pairs can be defined within a triad-connection formulation. This choice is primarily motivated by its advantages in formulations that connect to gauge theory and/or allow for the inclusion of fermions. Besides, by retaining this set of variables we maintain consistency with the notation of Refs.~\cite{MenaMarugan:2024qnj, MenaMarugan:2025anx} on the Hamiltonian treatment of metric perturbations. The relation between this formulation and the metric formulation, usually considered when studying (nonrotating) BH perturbations~\cite{Martel:2005ir, Lenzi:2021wpc} will be introduced in the next Subsection. Therefore, instead of using the standard canonical pair for Hamiltonian analysis, consisting of the background spatial metric two-form $g^{(3)}$ (or, equivalently, the associated tensor $g^{(3)}_{ij}$) and its conjugate momentum $\Pi^{(3)}$, we adopt the densitized triad $E$ and an associated $su(2)$-connection $A$. This change does not affect the physical content of the discussion. The additional degrees of freedom contained in the triad and connection are not physical, but constrained by the gauge $SU(2)$ invariance of the new formalism. In fact, for the triad-connection variables we have the following expression for the spatial metric:
\begin{equation}
    g^{(3)}= \frac{\delta(E,E)}{\text{det}(E)} = \frac{p_b^2}{L_o^2|p_c|} \text{d}\zeta^2 + |p_c|(\text{d}\theta^2 + \sin^2\theta\text{d}\phi^2)\, ,
\end{equation}
where the radial coordinate (of the interior Kantowski-Sachs geometry) is denoted by $\zeta$, and the usual angular coordinates by $\theta$ and $\phi$. The time coordinate (of the interior geometry), although not explicitly present, is denoted by $\chi$. The functions $p_b$ and $p_c$, known as densitized triad variables, vary with $\chi$ and determine $E$. Finally, $L_o$ is a fiducial length introduced to compactify the spatial sections and to avoid infrared divergences related with infinite volumes. Under careful analysis, $L_o$ can be taken as arbitrarily large, allowing one to recover the noncompact limit.

A similar correspondence can also be established between the connection variables and the conjugate momenta of the metric, $\Pi^{(3)}$, leading to
\begin{equation}
    \Pi^{(3)} = -2|p_c|\frac{L_o^2}{p_b^2}\Omega_b\sin\theta\partial_\zeta^2 - \frac{1}{|p_c|}(\Omega_b + \Omega_c) \left(\sin\theta\partial_{\theta}^2 + \csc\theta\partial_{\phi}^2\right)\,,
\end{equation}
where $\Omega_b = bp_b/L_o$ and $\Omega_c = cp_c/L_o$, while $b$ and $c$, referred to as connection variables, vary with $\chi$ and determine $A$, which encodes the information about the extrinsic curvature. Since the background shift vector can be set to zero owing to the symmetry of the $\chi$-constant sections, the full spacetime metric is recovered by including the lapse function $N$ in the metric expression. In our case, we adopt a densitized lapse $\tilde{N}$ that simplifies the Hamiltonian. At the practical level, this choice does not affect the classical physical results, as it is merely related to the choice of parametrization in the dynamical direction. The relation between the standard and densitized lapse functions is given by $N^2(\chi) = |p_c|p_b^2\tilde{N}^2(\chi)/L_o^2$. Under these considerations, the full spacetime metric takes the form
\begin{equation}
    g^{(4)} = -|p_c|\frac{p_b^2}{L_o^2} \tilde{N}^2 \text{d}\chi^2 + \frac{p_b^2}{L_o^2|p_c|} \text{d}\zeta^2 + |p_c|(\text{d}\theta^2 + \sin^2\theta\text{d}\phi^2)\,.
\label{KS-metric}
\end{equation}

The spacetime symmetries play a crucial role in the construction of the Hamiltonian formulation. Thanks to these symmetries, our model is significantly simplified. Starting from the standard expression for the ADM action and substituting the expressions for $g^{(3)}$ and $\Pi^{(3)}$ corresponding to the considered class of backgrounds, the action (after performing a spatial integration on our compactified sections\footnote{In all our analysis, we neglect any possible initial and final surface contributions.}) takes the following form:
\begin{equation}
    S_0 = \int \bigg( p_b\text{d}b + \frac{1}{2}p_c\text{d}c - \tilde{H}_{\text{KS}}[\tilde{N}]\text{d} \chi\bigg)\,.
\end{equation}
This model represents a fully constrained system. The first two terms of the action determine the phase space (pre)symplectic structure, while the last term constrains the relation between canonical variables. This means that, unlike in an unconstrained Hamiltonian formulation, not all trajectories in phase space are physical. Only those lying on the (hyper)surface generated by the constraint are considered physical. Dynamics follow from combining the Poisson bracket structure and the zero order Hamiltonian constraint, which are given by
\begin{equation}
    \{b,p_b\}_{\text{B}} = 1\,, \quad \{c,p_c\}_{\text{B}} = 2\,, \quad \tilde{H}_{\text{KS}}[\tilde{N}] = -\tilde{N}\frac{L_o}{2}\left( \Omega_b^2 + \frac{p_b^2}{L_o^2} + 2\Omega_b\Omega_c \right)\, .
\end{equation}
Here, the subscript $\text{B}$ indicates that the Poisson brackets refer to background variables. With these two ingredients, we have fully characterized the background Hamiltonian formulation. It is important to emphasize that a nonvanishing Hamiltonian is not required. In practice, to generate dynamics we use the expression above for the Hamiltonian constraint when performing computations, and only afterward do we impose the condition $\tilde{H}_{\text{KS}} = 0$, which corresponds to working on the constraint surface of the phase space.

\subsection{Triad-connection variables for the black hole geometry}

Within the context of BH physics, the Kantowski-Sachs spacetime can describe the interior region of a nonrotating BH, namely a Schwarzschild spacetime if we restrict the discussion to classical solutions to GR. In this work, we would like to highlight that, from the Hamiltonian point of view, the interior and exterior geometries can be treated within exactly the same formalism, with due considerations on the physical interpretation. This idea has motivated our choice of notation $(\chi,\zeta)$ for the nonangular coordinates, intended to minimize the possible confusion arising from their interpretation as temporal and spatial coordinates, and focus instead on the phase space structure. Indeed, owing to the spherical symmetry of the problem and to staticity (or, alternatively, hypersurface orthogonality of the additional Killing symmetry), the relevant dynamics is effectively encoded in a reduced two-dimensional submanifold coordinated by $\chi$ and $\zeta$. This geometrical picture admits in principle an associated canonical formulation\footnote{The well posedness of this formulation may break down in a strong singularity. We restrict our discussion to backgrounds or regions of them without such singularities.} with a well-defined phase space, a symplectic structure, and a constraint. Within the phase space, one can work with either $(b,c,p_b,p_c)$ or with the variables obtained by the following complex canonical transformation:
\begin{equation}
    b\rightarrow -ib\,,\quad p_b \rightarrow i p_b\,,
    \label{KS-to-Schw}
\end{equation}
so that the sign in the conformal factor of the set of spherically symmetric orbits is flipped, and so is the role and timelike/spacelike character of the coordinate variables $(\chi,\zeta)$. The metric~\eqref{KS-metric} becomes
\begin{equation}
    g^{(4)} = -\frac{p_b^2}{L_o^2|p_c|} \text{d}\zeta^2+|p_c|\frac{p_b^2}{L_o^2} \tilde{N}^2 \text{d}\chi^2  + |p_c|(\text{d}\theta^2 + \sin^2\theta\text{d}\phi^2)\,.
\end{equation}
In the Hamiltonian formulation, any phase space function has an associated Hamiltonian flow that dictates an evolution. Therefore, when working with the initial phase space variables, the evolution parameter associated with the Hamiltonian constraint can be interpreted as a time. Conversely, when performing the complex canonical transformation, the evolution parameter corresponding to the new real canonical set is instead interpreted as a radial parameter. Thus, although the formalism is the same in both cases, the physical interpretation is entirely different. We emphasize that neither interpretation derives from the other, rather, both provide equally valid descriptions from the Hamiltonian perspective. While they may appear to represent completely distinct scenarios, within this framework they are simply two sides of the same coin.

The Schwarzschild solution is then the solution of the background equations of motion in GR, so that the metric components satisfy the following conditions:
\begin{equation}
    \tilde{N} = -\frac{L_o^2}{p_b^2}= \frac{1}{\chi^2 f(\chi)}\,, \quad \frac{p_b^2}{L_o^2|p_c|} = - f(\chi)\,, \quad |p_c| = \chi^2\,,\quad \Omega_b = \chi f(\chi)\,,\quad \Omega_c = M\,,
    \label{schwarzschild-tetrad}
\end{equation}
where
\begin{equation}
f(\chi)\equiv \left(1-\frac{2M}{\chi}\right)\,,
\end{equation}
is the standard Schwarzschild function. It is then clear that, under the transformation~\eqref{KS-to-Schw}, the only effect in the metric is, in practice, the change from positive to negative values in the range of the Schwarzschild metric function $f$. The metric reads,
\begin{equation}
  g^{(4)} =  \frac{\text{d}\chi^2}{f(\chi)}- f(\chi) \text{d}\zeta^2  + \chi^2(\text{d}\theta^2 + \sin^2\theta\text{d}\phi^2)\,,
\end{equation}
where $f<0$ in the Schwarzschild interior, i.e. $\chi < 2M$, and $\chi$ is a time coordinate, while $f>0$ in the Schwarzschild exterior, i.e. $\chi > 2M$, and $\chi$ is a radial coordinate. This allows us to interpret $M$ as the ADM mass of the exterior BH spacetime.

To summarize, we argue that it may be more insightful to think of the phase space of the triad-connection variables, together with the Hamiltonian constraint, as the most fundamental tools for addressing the problem. Owing to the spherical symmetry of the spacetime, the relevant dynamics is effectively encoded in a reduced two-dimensional submanifold, described, in our case, by the coordinates $\chi$ and $\zeta$. If we momentarily set aside their physical interpretation and focus solely on the Hamiltonian framework, the picture that emerges is the following: we have a phase space characterized by $(b,c,p_b,p_c)$, with dynamics governed by the associated Hamiltonian constraint. Thus, we have a formulation for a set of variables that evolve along the flow generated by the Hamiltonian, but the physical interpretation of this evolution may not correspond to physical time in a strict sense. The interpretation of this evolution depends on the domain of definition of the variables. If we assume that the $b$-sector is real, then the physical picture, when connected to BH physics, corresponds to an interior description of the BH with timelike evolution. Conversely, if we consider a complexified (imaginary) $b$-sector, the same formalism yields a description of the exterior region, now interpreted as a radial evolution. This choice allows us to provide a unified Hamiltonian formulation for both the exterior and interior geometries. Later, when trying to relate the results in Kantowski-Sachs spacetime to BH physics, this perspective will be essential for bridging both interpretations and show that they share underlying symmetry structures.

\section{Perturbative Action}
\label{S:Perturbative-action}

We are now interested in studying first order perturbations over the previously described background. This was done in Refs.~\cite{MenaMarugan:2024qnj, MenaMarugan:2025anx} and we here simply review the Hamiltonian approach to determine gauge invariants for gravitational perturbations over a spherically symmetric spacetime. In doing so, we will concentrate our discussion on axial perturbations. We will also provide a dictionary between the typical notation from Refs.~\cite{Martel:2005ir, Lenzi:2021wpc} and the one of Ref.~\cite{MenaMarugan:2024qnj}.

The Hamiltonian framework is particularly insightful since it provides a natural separation into dynamical and gauge variables. Roughly speaking, the dynamical variables appear in the Hamiltonian while the gauge ones either do not appear in any of the constraints (including the Hamiltonian one) or, at most, become Lagrange multipliers in front of the constraints, which in turn provide the gauge transformation of the dynamical variables. Given the compactness of our sections, the perturbative action at second order, found by perturbing the ADM Hamiltonian form of the action, only contains quadratic terms in the first order perturbations if, as we will do, we treat all zero modes exactly at our perturbative truncation order. We refer to Refs.~\cite{Moncrief:1974vm,BrizuelaPhD, Brizuela:2006ne,Brizuela:2007zza,Brizuela:2008sk} for details and explicit expressions. More concretely, the discussed perturbative action takes the form
\begin{equation}
\triangle^{2}_{1}[S_0]= \frac{2}{\kappa} \int_{\mathbb{R}}\text{d}\chi\int_{\Sigma}\text{d}^3 x \left(p^{ij}\partial_{\chi}{h}_{ij} - C \triangle[\mathcal{H}] - B^i \triangle[\mathcal{H}_i]-\frac{N}{2}\triangle^{2}_{1}[\mathcal{H}] \right)\,,
\label{action}
\end{equation}
where $\kappa=16\pi$, $\Sigma=S_o^1 \times S^2$ is the compactified section, product of a circle and a 2-sphere, $\triangle^{2}_{1}$ denotes second order quantities that are composed of first order perturbations, $C = \triangle N$ is the first order variation of the lapse function, $B^i  = \triangle N^i$ is the first order variation of the shift vector, which vanishes in our background metric, $\mathcal{H}$ and $\mathcal{H}_i$ are the Hamiltonian and momentum constraints (that altogether generate spacetime diffeomorphisms), and $p^{ij}$ are the momenta associated to the induced three-metric perturbations $h_{ij}$ (corresponding to the perturbations of the metric $g^{(3)}_{ij}$).  

\subsection{Mode expansion of the perturbative action}

The following step is to exploit the symmetry of the background. Owing to the spherical symmetry, we can express every quantity in the perturbative action in (scalar, vector, and tensor) spherical harmonics for each parity, together with a Fourier decomposition of the $\zeta$ dependence on the circle. Note that while it is more common to expand in complex harmonics (see e.g. Refs.~\cite{Martel:2005ir, Lenzi:2021wpc, Lenzi:2024tgk,BrizuelaPhD}), the approach of Refs.~\cite{MenaMarugan:2024qnj, MenaMarugan:2025anx} uses real harmonics, in part to avoid introducing reality conditions on the perturbations that could obscure the effect of the complexification of the background variables. The harmonics expansion is particularly useful owing to the spherical symmetry of the background as it allows a neat separation of the equations for each harmonic number and parity.~\footnote{The harmonic numbers are usually named as $\ell \in \mathbb{N}_0$ and $m=-\ell,...,\ell$, with $\mathbb{N}_0$ indicating the natural numbers including zero. Parity of each harmonic is defined by how they transform under parity transformations, i.e. $(\theta, \phi)\rightarrow (\pi-\theta,\phi+\theta)$ (see e.g. Ref.~\cite{Lenzi:2021wpc}).} Despite choosing a real harmonics expansion, since we mostly follow the Hamiltonian treatment of Refs.~\cite{MenaMarugan:2024qnj, MenaMarugan:2025anx}, we will try to make clear contact between this and the formalism of Refs.~\cite{Martel:2005ir, Lenzi:2021wpc}. 

Therefore, we define the harmonic expansion of the metric in odd/axial parity modes [identified with the label ${\rm (ax)}$] and even/polar parity modes [identified with the label ${\rm (po)}$] as follows:
\begin{equation}
h_{ij}= \sum_{\ell m}\left( {}^{ \rm (ax)}h^{\ell m}_{ij}+{}^{ \rm (po)} h^{\ell m}_{ij}\right)\,.
\end{equation}
Only $\ell\geq 2$ multipoles will be considered in what follows. In fact, the $\ell=1$ and $\ell=0$ (with $n\neq 0$) modes correspond to pure gauge degrees of freedom, while the case $n=\ell=m=0$, also known as the zero mode, is already accounted for in the background dynamics up to the second perturbative order in our treatment. More details about these statements can be found in Refs.~\cite{Martel:2005ir, MenaMarugan:2024qnj, MenaMarugan:2025anx} (including the analysis of the finite number of modes with $n=0$ and $\ell=1$ \cite{MenaMarugan:2024qnj}). 

A similar motivation to the one that justifies the use of real harmonics holds for the (compactified) direction coordinated by $\zeta$, which is expanded in real Fourier modes. This introduces a discrete set of frequencies (see Refs.~\cite{MenaMarugan:2024qnj, MenaMarugan:2025anx}), as opposed to the case of an exterior, nonrotating BH spacetime, where one usually considers a Fourier expansion in the time domain that leads to a continuous spectrum of frequencies. The continuous frequency limit can be properly taken at the end of our calculations, once all the Fourier modes are included by suitably taking $L_o$ sufficiently large. The elements of the real Fourier basis, with period $L_o$, are
\begin{equation}
Q_{n,+}(\zeta)=\sqrt{\frac{2}{L_o}}\cos{(\omega_n \zeta)}\,,\quad Q_{n,-}(\zeta)=\sqrt{\frac{2}{L_o}}\sin{(\omega_n \zeta)}\,,
\end{equation}
for $n \geq 1$, and $Q_0(\zeta) = {1}/{\sqrt{L_o}}$ for $n=0$, where $\omega_{n} =2\pi n/L_o$. The metric perturbations are then also Fourier expanded on this real discrete mode as
\begin{equation}
h_{ij}= \sum_{\mathfrak{n},\lambda}\left( {}^{ \rm (ax)}h^{\mathfrak{n},\lambda}_{ij}+{}^{ \rm (po)}h^{\mathfrak{n},\lambda}_{ij} \right) Q_{n,\lambda} \,,\quad {}^{ \rm (ax/po)}h^{\mathfrak{n},\lambda}_{ij} =\int_{S^1_0} \text{d}\zeta\, {}^{ \rm (ax/po)}h^{\ell m}_{ij}\,Q_{n,\lambda} \,,
\end{equation}
where $\mathfrak{n}=(n,\ell,m)$, with $n=0,1,2,3...$, and $\lambda=+,-$ (and $\lambda$ a spurious label if $n=0$). The derivative of a function with respect to $\zeta$ has the following Fourier expansion:
\begin{equation}
\partial_\zeta F= \sum_{\mathfrak{n},\lambda} \lambda\omega_n F^{\mathfrak{n},-\lambda} Q_{n,\lambda}\,.
\label{Fourier-inversion}
\end{equation}
This will be useful later, especially in Sec.~\ref{Ss:RW-equation}.

The perturbations of the momenta $p_{ij}$ (with indices lowered with the three metric), the lapse, and the shift are also expanded accordingly,
\begin{eqnarray}
\begin{aligned}
p_{ij} &= \sqrt{\text{det}(g^{(3)})}\sum_{\mathfrak{n},\lambda}\left({}^{ \rm (ax)}p^{\mathfrak{n},\lambda}_{ij}+{}^{ \rm (po)}p^{\mathfrak{n},\lambda}_{ij}\right) Q_{n,\lambda}
\,, \\
C &= \sum_{\mathfrak{n},\lambda}\left({}^{ \rm (ax)}C^{\mathfrak{n},\lambda}+{}^{ \rm (po)}C^{\mathfrak{n},\lambda}\right) Q_{n,\lambda}\,, \quad 
B_{i} = \sum_{\mathfrak{n},\lambda}\left({}^{ \rm (ax)}B_{i}^{\mathfrak{n},\lambda}+{}^{ \rm (po)}B_{i}^{\mathfrak{n},\lambda}\right)  Q_{n,\lambda}\,.
\end{aligned}
\end{eqnarray}
Although we see that axial and polar perturbations can be handled in similar ways, the polar case is much more intrincate. For simplicity in the rest of our analysis, and concreteness in our results, from now on we will exclusively focus on the odd/axial part of the perturbations. As a result, we recognize harmonic coefficients of the spatial metric as
\begin{equation}
{}^{\rm (ax)}h^{\mathfrak{n},\lambda}_{i j}  = \begin{pmatrix} 0 & h_\zeta^{\mathfrak{n},\lambda}X^{\ell m}_{\theta} & h_\zeta^{\mathfrak{n},\lambda}X^{\ell m}_{\phi} \\[2mm]
h_\zeta^{\mathfrak{n},\lambda}X^{\ell m}_{\theta} & \multicolumn{2}{c}{\multirow{2}{*}{$h_2^{\mathfrak{n},\lambda} X_{AB}^{\ell m}$}} \\[2mm]
h_\zeta^{\mathfrak{n},\lambda}X^{\ell m}_{\phi} & &
\end{pmatrix}  
\,,
\label{harmonic-metric-perturbation}
\end{equation}
where we use the same notation as in Refs.~\cite{Martel:2005ir,Lenzi:2021wpc, MenaMarugan:2024qnj} for the different types of real vector and tensor axial harmonics.
The corresponding harmonic coefficients for the momenta, again focusing only on the axial sector, read
\begin{equation}
{}^{\rm (ax)}p^{\mathfrak{n},\lambda}_{i j}  = \begin{pmatrix} 0 & \tilde{p}_1^{\mathfrak{n},\lambda}X^{\ell m}_{\theta} & \tilde{p}_1^{\mathfrak{n},\lambda}X^{\ell m}_{\phi} \\[2mm]
\tilde{p}_1^{\mathfrak{n},\lambda}X^{\ell m}_{\theta} & \multicolumn{2}{c}{\multirow{2}{*}{$\tilde{p}^{\mathfrak{n},\lambda}_2 X_{AB}^{\ell m}$}} \\[2mm]
\tilde{p}_1^{\mathfrak{n},\lambda}X^{\ell m}_{\phi} & &
\end{pmatrix}
\,,
\end{equation}
where the tilde is introduced for later notational convenience (see Eq.~\eqref{sympl-transf}), while for the axial components of the lapse and shift we have
\begin{align}
{}^{\rm (ax)}C^{\mathfrak{n},\lambda} =0\,,\quad 
{}^{\rm (ax)}B_{i}^{\mathfrak{n},\lambda}= \left(0, h_\chi^{\mathfrak{n},\lambda} X_{A}^{\ell m} \right) \,.
\end{align}

A first background and mode dependent redefinition of the momentum variables is useful to simplify the action and reach genuine canonical commutation relations among configuration and momentum variables. Such redefinition is 
\begin{align}
p^{\mathfrak{n},\lambda}_{1} &= \frac{2 L^2_{o}}{p^2_{b}}\ell(\ell +1) \tilde{p}^{\mathfrak{n},\lambda}_{1}\,,\quad 
p^{\mathfrak{n},\lambda}_{2} =\frac{1}{2 p^2_{c}}\frac{(\ell +2)!}{(\ell -2)!} \tilde{p}^{\mathfrak{n},\lambda}_{2}
\,.
\label{sympl-transf}
\end{align}
After inserting the harmonic expansion and exploiting the compactness of the $\zeta$ direction, so that one can integrate the products of discrete Fourier modes over $S^1_0$, the second order axial action reduces to
\begin{equation}
{}^{\rm (ax)}\triangle^{2}_{1}[S_0]\bigg\vert_{\ell\geq 2}\!\!= \frac{2}{\kappa} \int_{\mathbb{R}} \text{d}\chi\left[\sum_{\mathfrak{n},\lambda}\big(p^{\mathfrak{n},\lambda}_{1} \partial_{\chi}{q}^{\mathfrak{n},\lambda}_{1}+p^{\mathfrak{n},\lambda}_{2}\partial_{\chi}{q}^{\mathfrak{n},\lambda}_{2}\big)  -\kappa\left( \mathcal{C}_{0}[h^{\mathfrak{n},\lambda}_{0}]+ H_{\rm ax}[\tilde{N}]\right)\right]\,,
\label{odd-action-symmred}
\end{equation}
where we have defined configuration variables and rescaled the lapse and shift perturbations as follows
\begin{equation}
q_1^{\mathfrak{n},\lambda} = -h_\zeta^{\mathfrak{n},\lambda}\,,\quad q_2^{\mathfrak{n},\lambda} =h_2^{\mathfrak{n},\lambda}\,,\quad h_0^{\mathfrak{n},\lambda} = -\frac{h_\chi^{\mathfrak{n},\lambda}}{\kappa}\,.
\label{variables-redef}
\end{equation} 
These simple redefinitions are again introduced to simplify the expressions. Here, $\mathcal{C}_0$ is the spatial integral of the axial part of $B^i \triangle[\mathcal{H}_i]$ (summed over $i=1,2,3$), expanded in harmonics and Fourier modes. The Hamiltonian $H_{\rm ax}[\tilde{N}]$ is the sum of all the contributions of the axial modes to the global Hamiltonian constraint coming from $N\triangle^{2}_{1}[\mathcal{H}]/2 $. To keep the discussion concise, we do not present the full expressions for the perturbative constraints and the Hamiltonian at this point. Instead, we refer the reader to Appendix \ref{App: iH}. The detailed derivation of these expressions can be found in Ref.~\cite{MenaMarugan:2024qnj}. 

In contrast to the background formulation (which corresponds to a fully constrained system by its own when unperturbed), at leading order in perturbation theory we encounter both perturbative constraints and a nonvanishing on-shell expression for the perturbative contribution to the Hamiltonian. As a brief overview, for each mode and considering the two polarities, there exist four perturbative gauge constraints (one axial and three polar) which, in the Hamiltonian approach, generate the corresponding perturbative gauge transformations at leading, first order \cite{MenaMarugan:2024qnj, MenaMarugan:2025anx}. These constraints are directly related to the additional gauge freedom introduced when perturbations are considered and can be understood as the linearized perturbative version of the Hamiltonian and momentum constraints of GR.~\footnote{A key advantage of the Hamiltonian approach (beyond facilitating the counting of physical degrees of freedom) is that it provides a simple test for gauge invariance: a quantity is perturbatively gauge invariant if it has vanishing Poisson brackets with the perturbative constraints of the system.} In addition, we have a nonvanishing (even when the perturbative constraints are imposed) perturbative contribution to the Hamiltonian. At our truncation order, axial and polar perturbations remain decoupled and can be analyzed independently. 

\subsection{Perturbative gauge invariants}
\label{Ss:perturbative-GI}

In the previous Subsection we obtained the desired separation, exploiting the spherical symmetry, among dynamical variables, perturbative gauge constraints, and Lagrange multipliers. However, we can still refine the expression of the action by writing the Hamiltonian in terms of gauge invariant quantities, the relevant ones for any physical phenomena. To do so, we will perform a canonical transformation on our perturbative variables, taking the background as fixed. Notably, the canonical transformation can be extended to the background if it is considered dynamical by correcting its variables with quadratic contributions of the perturbations. The fact that these corrections are quadratic is required by the consideration of zero modes to describe the background. These corrections are compatible with our perturbative order of truncation in the action, and can be consulted in Ref.~\cite{MenaMarugan:2024qnj}. Restricting our attention to the first order perturbations of the metric, we thus consider transformations of the type
\begin{equation}
    \{q_I^{\mathfrak{n},\lambda},p_{I'}^{\mathfrak{n}',\lambda'}\}_{\text{P}} = \kappa\delta_{II'}\delta^{\mathfrak{n}\mathfrak{n}'}\delta^{\lambda\lambda'}\, \quad \xrightarrow{\text{Canon. Trans.}} \quad \{Q_I^{\mathfrak{n},\lambda},P_{I'}^{\mathfrak{n}',\lambda'}\}_{\text{P}} = \kappa\delta_{II'}\delta^{\mathfrak{n}\mathfrak{n}'}\delta^{\lambda\lambda'}\,.
\end{equation}
Here, the subscript $\text{P}$ indicates that the Poisson brackets refer to perturbative variables. Recall that the index $I$ runs from $1$ to $2$ for our axial perturbations. The key requirement on the desired transformation for these perturbations is that one of the new variables $\{ Q_I^{\mathfrak{n},\lambda}, P_I^{\mathfrak{n},\lambda}\}$ has to be proportional to the axial perturbative constraint. Another variable is then found under the requirement that the two commute. The transformation is completed by finding canonical pairs to the two previous variables. In the axial case, the perturbative constraint responsible of generating gauge transformations has the following general form [see Eq.~\eqref{constraint-C0}]:
\begin{equation}
\mathcal{C}_0 = \sum_{\mathfrak{n},\lambda} h_0^{\mathfrak{n},\lambda}F[q_{I}^{\mathfrak{n},\lambda},p_{I}^{\mathfrak{n},\lambda}]
\,.
\end{equation}
Then the starting point for finding the canonical transformation is the requirement that, e.g. $P_2^{\mathfrak{n},\lambda} \propto F[q_I^{\mathfrak{n},\lambda},p_{I}^{\mathfrak{n},\lambda}]$. Any $Q_1^{\mathfrak{n},\lambda}$ commuting with $P_2^{\mathfrak{n},\lambda}$, i.e. $\{ Q_1^{\mathfrak{n},\lambda} , P_2^{\mathfrak{n},\lambda} \}=0$, will be automatically gauge invariant while $Q_2^{\mathfrak{n},\lambda} $, which is found for each mode by requiring that its Poisson brackets with $ P_2^{\mathfrak{n},\lambda} $ are equal to one, i.e. $\{ Q_2^{\mathfrak{n},\lambda} , P_2^{\mathfrak{n},\lambda} \}=1$, will not be gauge invariant but pure gauge. Determining then the other canonical momentum $P_1^{\mathfrak{n},\lambda}$, the symplectic structure of the phase space of the axial perturbations remains unchanged, since the transformation is canonical. 

Moreover, a redefinition of the Lagrange multipliers with perturbative correction terms, compatible within the considered perturbative order, is introduced to restructure the action so that irrelevant terms proportional to $P_2^{\mathfrak{n},\lambda}$, which therefore vanish on shell, are not present in the Hamiltonian. The axial perturbative constraint then becomes
\begin{equation}
    \label{liga}
    \mathcal{C}_0[\tilde{h}_0^{\mathfrak{n},\lambda}] = -\sum_{\mathfrak{n},\lambda} 2\tilde{h}_0^{\mathfrak{n},\lambda}P_2^{\mathfrak{n},\lambda}\,, 
\end{equation}
where $\tilde{h}_0^{\mathfrak{n},\lambda}$ represents the axial contribution to the linear perturbation of the shift vector $h_0^{\mathfrak{n},\lambda}$, along with aforementioned perturbative correction terms included solely to ensure consistency in the overall expression for the perturbative action. Since $h_0^{\mathfrak{n},\lambda}$ has no physical relevance, these additional terms are not pertinent to the current discussion. As a result of all these operations, the nonvanishing contribution to the Hamiltonian takes the form
\begin{equation}
    \label{Hamiltonian-QP}
         \kappa \hat{H}_{{\rm ax}}[\tilde{N}]=  \sum_{\mathfrak{n},\lambda}\frac{\tilde{N}}{2}\left(A_{{\rm ax}}[Q_1^{\mathfrak{n},\lambda}]^2 + B_{{\rm ax}}[P_1^{\mathfrak{n},\lambda}]^2 + C_{{\rm ax}}Q_1^{\mathfrak{n},\lambda}P_1^{\mathfrak{n},\lambda}\right)\,.
\end{equation}
The three coefficients appearing in the expression of this Hamiltonian are explicitly given in Appendix \ref{App: HC}. 

From the above expressions, we can deduce that $(Q_1^{\mathfrak{n},\lambda},P_1^{\mathfrak{n},\lambda})$ represent the first order perturbative gauge invariant pair which constitute the physical perturbative degrees of freedom. The relevant dynamics (namely, the equations of motion for the invariant pair) can be derived directly from the obtained Hamiltonian. To conclude the discussion, we present the explicit expression for the effective action governing the perturbations. It resembles the form of the background action but includes additional terms coming from the perturbative degrees of freedom. The axial perturbative action is given by
\begin{equation}
{}^{\rm (ax)}\Delta^2_1[S] = \frac{2}{\kappa} \int_{\mathbb{R}} \text{d}\chi\left[\sum_{\mathfrak{n},\lambda}\big( P_1^{\mathfrak{n},\lambda}\partial_{\chi}{Q}_1^{\mathfrak{n},\lambda}+P_2^{\mathfrak{n},\lambda}\partial_{\chi}{Q}_2^{\mathfrak{n},\lambda}\big) - \kappa\left(\mathcal{C}_0[\tilde{h}_0^{\mathfrak{n},\lambda}] +  \hat{H}_{{\rm ax}}[\hat{N}]\right)\right]\,,
\end{equation}
where $\hat{N}$ is the standard densitized lapse function $\tilde{N}$ plus perturbative corrections, which can be found in Ref.~\cite{MenaMarugan:2024qnj}. These expressions will serve as the starting point for our perturbative analysis of the first order gauge invariants. Combined with the background dynamics and the insights provided on the role of canonical transformations, this framework will allow us to develop the analysis of DTs. 

\section{Darboux covariance and canonical transformations}
\label{S:darboux-symmetry}

In this Section, we lay the general structure that we will use to search for master functions starting from Eq.~\eqref{Hamiltonian-QP}. We first introduce the general features of the canonical transformation leading to a diagonal axial Hamiltonian, i.e. without crossed terms of the form $QP$. We then briefly review the Darboux covariance picture of Refs.~\cite{Lenzi:2021wpc,Lenzi:2021njy,Lenzi:2024tgk}. After imposing further restrictions on the general diagonalizing transformation, we show that the targeted diagonal Hamiltonians are actually related by what we call generalized DTs, clarifying how these can be described as canonical transformations. We stress that transformations of this type have a purely formal role at this level and that they can have nontrivial physical implications and introduce nonlocality features. In fact, demanding a diagonal Hamiltonian is simply a restriction on the canonical momentum chosen in each gauge invariant pair, and it is just in this sense that the diagonal requirement removes ambiguities in our canonical transformation. A physically well-posed local dynamics for the gauge invariant perturbations should rather follow from the properties of the remaining terms in the diagonal quadratic Hamiltonian for such perturbations. However, the aforementioned generalized transformations provide a useful guiding principle to obtain canonical Hamiltonians for the master functions. In physical cases, the possible problems with nonlocality are excluded and clarified by restricting to (essentially) frequency independent potentials in the perturbative Hamiltonian.

\subsection{Diagonalizing canonical transformation}
\label{Ss:diagonalizing-canonical}

Let us consider a generic quadratic Hamiltonian $H_1$ generating evolution in the direction coordinated by $\chi$, with the same structure as in Eq.~\eqref{Hamiltonian-QP} and for a densitized lapse set equal to the unit, $\tilde{N}=1$. Explicitly,
\begin{equation}
    H_1 = \frac{1}{2}\left(A q^2 + B p^2 + C p q\right)\,,
    \label{H1}
\end{equation}
with corresponding equation of motion for the configuration variable $q$ given by 
\begin{equation}
{q}^{\prime\prime} - \frac{{B}^{\prime}}{B}{q}^{\prime}+Vq=0\,,\quad V = AB -\frac{C^2}{4} -\frac{B}{2}\left(\frac{C}{B}\right)^{\prime}
\,,
\label{H1bis}
\end{equation}
where from now on, except when stated otherwise, the prime denotes the derivative with respect to $\chi$ divided by the densitized lapse, which is one with our fixation of this lapse in this Section.

Let us start by defining a $\chi$-dependent\footnote{Later in our discussion, we will restrict to a dependence on $\chi$ that arises exclusively from a dependence on the background geometry.} linear canonical transformation as follows
\begin{equation}
\left\{
\begin{array}{l}
q = \alpha(\chi) Q +\beta(\chi) P\\ [2mm]
p = \gamma(\chi) Q +\delta(\chi) P
\end{array}\right. \,,\quad {\rm with} \quad
\alpha\delta-\beta\gamma=1\,,
\label{canonical}
\end{equation}
and otherwise free functions $\alpha$, $\beta$, $\gamma$, and $\delta$. Such canonical relations emerge e.g. from a type-$1$ $\chi$-dependent generating function associated with the transformation, given by
\begin{equation}
    F_{(1)}(q,Q) = \frac{\delta q^2 + \alpha Q^2}{2\beta} - \frac{qQ}{\beta}\,, \quad p = \frac{\partial F_{(1)}}{\partial q}\,, \quad P = -\frac{\partial F_{(1)}}{\partial Q}\,.
\end{equation}
In Hamiltonian systems, when a canonical transformation depends explicitly on the evolution parameter, the Hamiltonian expressed in terms of the new variables generally differs from the original Hamiltonian. In such cases, the generating function approach for describing canonical transformations becomes particularly useful, providing a more systematic framework than the standard formulation. It is worth noting that the type of generating function used [e.g., $F_{(1)}(q,Q)$, $F_{(2)}(q,P)$, etc.] simply indicates the set of variables on which the function depends; beyond that, the classification has no deeper significance. The primary advantage of using a generating function lies in the straightforward computation of the transformed Hamiltonian. Specifically, the new Hamiltonian is given by the original Hamiltonian plus a correction term involving the partial derivative, in the direction of the Hamiltonian evolution, of the generating function. Importantly, this correction term is independent of the type of generating function employed. As long as the generating functions describe the same canonical transformation, they yield the same correction, equal to the explicit derivative of the generating function with respect to $\chi$. On the other hand, an alternative procedure to calculate the new Hamiltonian is to start from the action of the system in Hamiltonian form and substitute in it the canonical change of variables. We note that, in the case with compact sections under consideration, there are no boundary contributions to the Hamiltonian in the action.~\footnote{It would be interesting to investigate if the Hamiltonian that arises in the noncompact limit requires boundary terms or suitable boundary conditions on the perturbations in order to be differentiable on the phase space of the corresponding system.}

Adopting any of these procedures, the resulting new Hamiltonian $H_2$ takes the quadratic form
\begin{equation}
    H_2 = \frac{1}{2}\left(\mathcal{A} Q^2 + \mathcal{B} P^2 + \mathcal{C} P Q\right)\,,
    \label{H2}
\end{equation}
where the new coefficients are related to the old ones by the following relations:
\begin{equation}
\begin{aligned}
\mathcal{A} &= A\alpha^2 +B\gamma^2 +C\alpha\gamma +\alpha {\gamma}^{\prime} -\gamma {\alpha}^{\prime}\,,\\
\mathcal{B} &= A\beta^2 + +B\delta^2 +C\beta\delta + \beta {\delta}^{\prime} -\delta {\beta}^{\prime}\,,\\
\mathcal{C} &= 2A\alpha\beta +2B\gamma\delta +C\left(\alpha\delta +\beta\gamma\right) +2\left(\alpha {\delta}^{\prime} -\gamma {\beta}^{\prime} \right)\,.
\end{aligned}
\label{constraints-ABC}
\end{equation}
These equations can be used to provide useful constraints on the canonical coefficients $\alpha$, $\beta$, $\gamma$, and $\delta$ under the following requirements on the final form of the Hamiltonian. a) We want the Hamiltonian to be diagonal, i.e. that there are no crossed terms proportional to $QP$. This condition allows us to simplify the quantization procedure while removing ambiguities in the choice of canonical momentum. Furthermore, this is a desirable physical requirement if one wants to avoid potential momentum instabilities under quantization of the system. Additionally, in the case of an infinite number of canonical pairs, crossed terms in the Hamiltonian typically prevent the unitary implementation of the associated evolution. b) We want the coefficient in front of the squared momentum to be constant, to guarantee that the momentum $P$ is simply the $\chi$ derivative of the configuration variable $Q$. This implies setting $\mathcal{B}=\mathrm{const}$, and removes the ambiguity of scaling the configuration variable by a $\chi$-dependent function. Under these two restrictions, the Hamiltonian and the equation of motion adopt, respectively, the following forms:
\begin{equation}
H_2 = \frac{1}{2}\left(\mathcal{A} Q^2 + \mathcal{B} P^2\right) \quad  {\rm with} \quad \mathcal{B}={\rm const}\, , \qquad {Q}^{\prime} = \{Q,H_2\} = \mathcal{B} P\,,\quad {P}^{\prime} = \{P,H_2\} = -\mathcal{A} Q \,.
\label{H-diagonal}
\end{equation}
Indeed, up to a constant scaling of the momentum allowed by $\mathcal{B}$, we are left with just a total potential $v$, which in this notation coincides with $-\mathcal{A}\mathcal{B}$. The evolution equation for the new configuration variable can be written in the form
\begin{equation}
{Q}^{\prime\prime} - v Q=0\,,\quad  v=-\mathcal{A}\mathcal{B}= \mathcal{B} \left ({w}^{\prime}+w^2 -\frac{B}{\beta^2}\right) \quad {\rm with} \quad w= - \frac{\alpha}{\beta}
\,.
\label{V-diagonal}
\end{equation}
This expression for the potential is obtained by solving the constraints $\mathcal{B}=\mathrm{const}$ and $\mathcal{C}=0$ [second and third lines in Eq.~\eqref{constraints-ABC}] for $A$ and $C$, and exploiting the condition ~\eqref{canonical} that the transformation be canonical. Therefore, the set of canonical transformations~\eqref{canonical} satisfying that $\mathcal{B}$ be constant and $\mathcal{C}$ vanish describes infinite possible Hamiltonians with potentials given by Eq.~\eqref{V-diagonal}. In the following subsections we investigate the canonical transformations among such Hamiltonians and their relation to DTs.

\subsection{Darboux covariance}
\label{Ss:Darboux-covariance}

We here briefly review what is meant by a DT in connection with the findings of Ref.~\cite{Lenzi:2021wpc, Lenzi:2021njy}, where it was shown that the decomposition of perturbative Einstein equations into master equations for the linearized degrees of freedom allows infinite physically equivalent choices, each defined by a pair composed of a potential and a master function, $(V, \Psi)$, all these pairs being related by DTs. Let us consider the following wave equation:
\begin{equation}
    \left[\square^{(2)} - \Theta\right]\Psi = 0 \,.
\end{equation}
This is the typical wave equation that the master functions satisfy, with due considerations on the sign of the potential, for (Schwarzschild) BH perturbations~\cite{Martel:2005ir,Lenzi:2021wpc} as well as Kantowski-Sachs spacetime perturbations (see Sec.~\ref{Ss:RW-equation} and Refs.~\cite{MenaMarugan:2024qnj, MenaMarugan:2025anx}). The Laplace-Beltrami operator $\square^{(2)}$ in the set of orbits of spherical symmetry is defined in Appendix~\ref{App: LB}. Using the coordinates $(\chi, \zeta)$ introduced before, we can write the wave equation as
\begin{equation}
    \left[-\partial^2_\sigma +\partial^2_\zeta + V\right]\Psi = 0 \,,
\end{equation}
where the derivative with respect to $\sigma$ is defined by $|p_c|\tilde{N} \partial_\sigma= \partial_\chi$, and $V$ is related to $\Theta$ by a multiplicative background factor (in fact, this factor is the two-dimensional conformal factor for the set of orbits of spherical symmetry, see Appendix~\ref{App: LB}). More concretely,
\begin{equation}
    V(\chi)= -\frac{p_b^2}{L_o^2 |p_c|}\Theta(\chi)\,.
    \label{V-Theta}
\end{equation}
Note that, owing to the change of sign in the conformal factor from the exterior to the interior region under Eq.~\eqref{KS-to-Schw}, the potential will also change sign accordingly. The overall minus sign convention is arbitrary and just chosen so that the signs of $V$ and $\Theta$ agree in the exterior. Irrespectively of whether we consider the exterior or interior spacetime, we can expand in Fourier modes with respect to the coordinate $\zeta$ and obtain
\begin{equation}
    \partial^2_\sigma \Psi +(\omega^2 - V)\Psi = 0 \,,\quad V=V(\sigma)\,.
    \label{master-frequency-ext-int}
\end{equation}
For simplicity, we obviate here the labels $(n,\lambda)$ of the Fourier mode and its frequency.

A DT relating two pairs, $(\Phi,v)$ and $(\Psi,V)$, is defined as follows:
\begin{eqnarray}
\left[-\partial^2_\sigma +\partial^2_\zeta + V\right] \Psi = 0
~\longrightarrow~
\left\{
\begin{array}{l}
\Phi =  \Psi^{}_{,\sigma}  + W\,\Psi 
\\[2mm]
v = V + 2\,W^{}_{,\sigma}
\\[2mm]
W^{}_{,\sigma}- W^2 + V = \mathbf{c} 
\end{array}
\right.
\!\!
\longrightarrow
~
\left[-\partial^2_\sigma +\partial^2_\zeta + v\right] \Phi = 0\,.
\label{DT}
\end{eqnarray}
Here $W$ is the DT generating function, which satisfies a Riccati equation with an arbitrary constant $\mathbf{c} $. Note that this Riccati equation can be equivalently rewritten as $W^{}_{,\sigma}+ W^2 + \mathbf{c} =v$ [compare with the second equation in Eq. \eqref{V-diagonal}]. 

In the exterior and when restricted to the Schwarzschild case, the physical consequence of this transformation becomes clear in the frequency domain, where all Darboux-related potentials can be shown to be isospectral~\cite{Lenzi:2021njy} under the standard black-hole condition that the modes be purely ingoing at the event horizon and purely outgoing at spatial infinity. While there does not seem to be a full proof about mode equality for ``any'' constant $\mathbf{c}$ for Schwarzschild spacetime, in Ref.~\cite{Yurov:2018ynn} it is demonstrated that there exist infinite such solutions. Moreover, in Ref.~\cite{Lenzi:2021njy} it has been shown that solutions to the Riccati equation with an arbitrary constant have the same Korteweg-de Vries integrals and in Refs.~\cite{Lenzi:2022wjv,Lenzi:2023inn} it was found that these integrals completely determine the graybody factors, all of these results referred to the Schwarzschild solution. Since the quasinormal modes are the poles of the graybody factors for this specific spacetime, then this provides at least an indirect proof in that case. Still in the Schwarzschild case, the isospectrality between axial and polar parity sectors, and not between Darboux related axial potentials, was already pointed out by Chandrasekhar~\cite{1980RSPSA.369..425C,Chandrasekhar:1992bo}, but it had first been realized to be consequence of a Darboux relation between the axial Regge-Wheeler potential and the polar Zerilli potential for such a spacetime in Ref.~\cite{Glampedakis:2017rar}. Again, the Korteweg-de Vries integrals provide clear and simple indicators of isospectrality breaking~\cite{Lenzi:2025kqs} when they differ between two potentials. The notion of isospectrality is therefore tied to the nature of the background and one cannot make completely general statements beyond the Schwarzschild solution. 

In Ref.~\cite{Lenzi:2021njy} (see also Ref.~\cite{Lenzi:2024tgk}) the full picture for gravitational perturbations of spherically symmetric BHs has been obtained and the role of DTs as symmetries of the equations has been shown. It has been found that, in each parity sector, there are actually infinite possible master equations describing the dynamics of the perturbations and that all of them are related by DTs. Therefore, there is an infinite tower of isospectral master equations for each parity while each parity sector is also isospectral to the other. This is the hidden symmetry that has been called Darboux covariance~\cite{Lenzi:2021njy}. The physical implications on the spectrum in the interior case, or even in the exterior for effective backgrounds that are not genuine vacuum solutions in GR, will have to be analyzed in detail in future works since the effective modifications may not preserve the asymptotic structure or subtly affect the standard black-hole conditions on the modes.

As a last remark, let us mention that the Riccati equation for the Darboux generating function $W$ allows us to rewrite Eq.~\eqref{master-frequency-ext-int} as
\begin{equation}
    \left(\partial_{\sigma}^2 +W^{}_{,\sigma}- W^2\right)\Psi = -\left(\omega^2 -\mathbf{c} \right)\Psi\,,
\end{equation}
irrespectively of the sign on the potential. This leads to realizing that the two Darboux related potentials correspond to supersymmetric partners, so that the master equation~\eqref{master-frequency-ext-int} can be factorized into
\begin{equation}
    \left(\partial_\sigma   - W   \right)\left(\partial_\sigma + W \right)\Psi = -(\omega^2 -\mathbf{c}) \Psi\,,\quad \left(\partial_\sigma  + W   \right)\left(\partial_\sigma - W \right)\Phi = -(\omega^2 - \mathbf{c}) \Phi\,.
\end{equation}
Note that the supersymmetric formulation does not distinguish at all between interior or exterior description.

\subsection{Darboux transformations in a canonical setting}
\label{Ss:darboux-canonical}

We can now investigate the relation existing among the Hamiltonians obtained after the diagonalization procedure described in Sec.~\ref{Ss:diagonalizing-canonical}. These are defined by Eq.~\eqref{H-diagonal}, i.e. they are diagonal and with constant coefficient in front of the squared momentum. In order to describe this family of Hamiltonians, let us start from the Hamiltonians in Eqs.~\eqref{H1} and~\eqref{H2} but with
\begin{equation}
    B = \mathcal{B} = 1\,, \quad C = \mathcal{C} = 0\,, \quad A=-V\,, \quad \mathcal{A}=-v\,,
    \label{H1-H2-diagonal}
\end{equation}
so that the terms $A$ and $\mathcal{A}$ act solely as potential terms, as in Eq.~\eqref{H-diagonal}. To simplify our discussion, we have taken here the constant value of $\mathcal{B}$ (and $B$) equal to one. One can always suitably scale the canonical momentum and perform an inverse scaling of the configuration variable to reintroduce any other constant value of $\mathcal{B}=B$ different from zero.  

Consider then the following linear combinations of the relations in Eq.~\eqref{constraints-ABC}:
\begin{equation}
    \begin{aligned}
        & \beta\frac{\mathcal{C}}{2} = \alpha \mathcal{B}  + {\beta}^{\prime} - \delta B - \beta\frac{C}{2}\,,\\
        &\alpha\frac{\mathcal{C}}{2} = \beta \mathcal{A} - {\alpha}^{\prime} + \gamma B + \alpha\frac{C}{2}\,,\\
        &\alpha^2 \mathcal{B}= \beta^2 \mathcal{A} + \alpha\beta C+ B(\alpha\delta+\beta\gamma) - (\alpha \beta)^{\prime}\,.
         \label{constraints-ABC-mix}
    \end{aligned}
\end{equation}
Imposing the conditions in Eq.~\eqref{H1-H2-diagonal} ,we can solve the first and second lines in Eq.~\eqref{constraints-ABC-mix}, respectively, for $\delta$ and $\gamma$ in terms of $\alpha$ and $\beta$, i.e. $\delta = \alpha + {\beta}^{\prime}$ from the first equation, and $\gamma = {\alpha}^{\prime}-\beta \mathcal{A}$ from the second equation. Therefore, the canonical transformation can be written as
\begin{equation}
\left\{
\begin{array}{l}
q= \alpha Q+\beta P\\ [2mm]
p = \left( {\alpha}^{\prime}-\beta \mathcal{A}\right) Q+ (\alpha + {\beta}^{\prime}) P
\end{array}\right. \quad.
\label{GDT-q}
\end{equation}
The third line in Eq.~\eqref{constraints-ABC-mix} becomes
\begin{equation}
{w}^{\prime} + w^2  -v = \frac{1}{\beta^2}\quad \mathrm{with}\quad v = V - \frac{2 {\alpha}^{\prime}}{\beta} - \frac{{\beta}^{\prime\prime}}{\beta}\,,\quad w= -\frac{\alpha}{\beta}\,.
\label{GDT-v}
\end{equation}
The only difference with Eq.~\eqref{V-diagonal} is the contribution from $B$ and the multiplicative factor $\mathcal{B}$, which are here set equal to one. 

The relation among the potentials, together with the Riccati equation for $w$ and the transformation between the configuration variables, i.e. Eqs.~\eqref{GDT-q} and~\eqref{GDT-v}, were named ``generalized Darboux transformations'' in Ref.~\cite{Glampedakis:2017rar}. The above analysis provides their interpretation as canonical transformations connecting diagonal Hamiltonians (with momenta equal to the derivative of the configuration variables in the direction of evolution). The reason for this name is their appearance at the beginning of the original work of Darboux~\cite{1999physics...8003D}. These types of transformations can be useful in some cases. For example, when studying perturbations of Kerr spacetime they are used to transform long range potentials into short range ones~\cite{Chandrasekhar:1976zz, Sasaki:1981kj}. They also allow one to obtain raising and lowering
operators for spin-weighted spheroidal harmonics~\cite{Shah:2015sva}. However, if one wants to preserve a local physical formalism, in general they are just formal transformations, essentially identifiable as the most general transformations that map, with first order differential operators, the second order equation~\eqref{V-diagonal} into another one of the same kind. Indeed, as noted in Ref.~\cite{Yurov:2018ynn}, these generalized DTs do not generically provide transformations between isospectral potentials even in Schwarzschild spacetime, a fact that indicates again their formal character for our purposes. As stressed in Ref.~\cite{Yurov:2018ynn}, the isospectral property is guaranteed in a Schwarzschild background only by a subclass of transformations: the DTs~\eqref{DT} described in Sec.~\ref{Ss:Darboux-covariance}, which are recovered by isolating the dependence on the squared Fourier frequency in the potentials, i.e. $V=V_0 -\omega^2$ and $v=v_0 -\omega^2$, and canceling the corresponding frequency term in Eq.~\eqref{GDT-v} by a suitable functional dependence on it of $\beta=\mathrm{const}$. If this dependence is not eliminated in the Riccati equation, the solution, and consequently the potentials, may depend on this frequency in a nontrivial way. In fact, it is straightforward to see from Eq. \eqref{GDT-v} that, if $\beta^{-2}-\omega^2$ is frequency dependent, the solution $w$ to the Riccati equation would also depend on the frequency, and hence (generically) all the coefficients
of the canonical transformation \eqref{GDT-q}.

In the case with constant $\beta$ when the frequency dependence can be eliminated, the discussed Riccati equation for the generalized DT reduces to
\begin{equation}
    {w}^{\prime} + w^2 - v = \frac{1}{\beta^2} \quad \mathrm{with} \quad  v = V + 2 {w}^{\prime}\,,\quad \beta = \mathrm{const}\,,
    \label{darboux-hamiltonian}
\end{equation}
which is consistent with Eq.~\eqref{DT} for $\mathbf{c}=-1/\beta^2$ and guarantees that $\delta=\alpha$, so that (using that now $P=Q^{\prime}$)
\begin{equation}
\left\{
\begin{array}{l}
q = \alpha Q+ \beta {Q}^{\prime}\\ [2mm]
p = \left( {\alpha}^{\prime}-\beta \mathcal{A}\right) Q+ \alpha {Q}^{\prime}\,.
\end{array}\right.
\end{equation}
While this set of transformations seems somewhat artificial at this stage, they will have a more natural implementation in Sec.~\ref{Ss:KS-master}.

Let us conclude this Section with a remark. Notice that if one considers the Riccati equation in Eq.~\eqref{darboux-hamiltonian}, it may look at first like the canonical transformation does not depend on the coefficients of the initial Hamiltonian. However, this conclusion based on the Riccati equation alone can be misleading, as the DT~\eqref{DT} requires (and the Hamiltonian analysis provides) both the relation among the two potentials and the Riccati equation for the superpotential. This means that Eq.~\eqref{darboux-hamiltonian} may be rewritten equivalently as
\begin{equation}
    {w}^{\prime} - w^2 + V = -\frac{1}{\beta^2} \quad \mathrm{with} \quad  v = V + 2 {w}^{\prime}\,.
    \label{darboux-hamiltonian-2}
\end{equation}
This is actually the starting point of the operational procedure to entirely fix the new Hamiltonian. Namely, I) solve the first two lines in Eq.~\eqref{constraints-ABC-mix} for $\delta$ and $\gamma$ in terms of $\alpha$ and $\beta$ as shown before; II) find the superpotential $w$ from the Riccati equation~\eqref{darboux-hamiltonian-2}, which depends on the starting potential $V$ and the suitable constant $\beta$, and then find the new potential $v$ (this fixes a relation between $\alpha$ and $\beta$); III) find the remaining coefficient of the transformation ($\alpha$ or $\beta$) through the condition that it be canonical; and IV) insert these expressions for the coefficients in the transformation to obtain the final result.

\section{Master variables and Darboux canonical symmetry}
\label{S:KS-Darboux}

In this Section, we follow the above general lines to introduce the subfamily of canonical transformations such that their effect on the Hamiltonian for the gauge invariant perturbations [see Eq.~\eqref{Hamiltonian-QP}] can be regarded as a mapping between Hamiltonians for the mode variables of master functions, with dynamics that reduce to a wave equation in two dimensions subject to a potential. The answer to this question has already been found in Refs.~\cite{Moncrief:1974vm} for the purely Schwarzschild case (in the axial sector) and in Refs.~\cite{MenaMarugan:2024qnj, MenaMarugan:2025anx} for Kantowski-Sachs cosmologies describing interior backgrounds. We here take a more general perspective which refines previous results and recovers the known master functions typically found in BH perturbation theory, while also showing how the hidden symmetry structure of Refs.~\cite{Lenzi:2021njy,Lenzi:2024tgk} emerges in the Hamiltonian framework. 

In summary, in Sec.~\ref{Ss:KS-master} we use the results from Sec.~\ref{S:darboux-symmetry} as a guide for the analysis of perturbations in the considered case of spherically symmetric backgrounds, so that the hidden Darboux structure becomes evident. In Sec.~\ref{Ss:RW-equation} we use the results of Ref.~\cite{new} to show that a particular solution corresponds to the Regge-Wheeler equation with the CPM master function. Moreover, by simply following the chain of canonical transformations, one has a systematic and efficient algorithm to reconstruct the metric perturbations from the canonical/master mode variables. Finally in Sec.~\ref{Ss:GS-master} we show the relation with the GS master function, which is actually a scale transformation of the configuration variables which implies a change in the associated evolution parameter. 

\subsection{Darboux related Hamiltonians in spherically symmetric backgrounds: Axial sector}
\label{Ss:KS-master}

Let us now follow the Hamiltonian diagonalization discussed in the previous Section, where now the starting Hamiltonian is the axial perturbative one given in Eq.~\eqref{Hamiltonian-QP}. Therefore, we analyze in the following exclusively axial gauge invariant perturbations and consider a canonical transformation of the form $\{ {}^{} Q_1^{\mathfrak{n},\lambda},{}^{}\! P_1^{\mathfrak{n},\lambda}\}\rightarrow \{ {}^{(1)} Q_1^{\mathfrak{n},\lambda},{}^{(1)}\! P_1^{\mathfrak{n},\lambda}\}  $.  Introducing the expressions of Appendix~\ref{App: HC} in Eq.~\eqref{constraints-ABC}, and obviating the mode labels to simplify the notation, we obtain the following coefficients\footnote{\label{backdyna}In the case that all the dynamical dependence of the coefficients of the considered canonical transformation comes from a background dependence, the derivatives with respect to $\chi$ can be computed with the background Hamiltonian as Poisson brackets in the phase space of the background variables. We do not make this fact explicit in the rest of our discussion in order to keep our notation simple.} for the new quadratic Hamiltonian of the perturbations:
\begin{align}
\label{constraint-A-KS}
A^{(1)}_{{\rm ax}} &= A^{}_{{\rm ax}}\alpha^2 +B^{}_{{\rm ax}}\gamma^2 +C^{}_{{\rm ax}}\alpha\gamma +\left(\alpha \gamma^{\prime} -\gamma \alpha^{\prime}\right) =\, |p_c| \left(\omega_n^2 -V_{\ell}^{{\rm ax}}\right) \,,\\
\label{constraint-B-KS}
B ^{(1)}_{{\rm ax}}&= A^{}_{{\rm ax}}\beta^2  +B^{}_{{\rm ax}}\delta^2 +C^{}_{{\rm ax}}\beta\delta + \left(\beta \delta^{\prime} -\delta \beta^{\prime}\right)=\, |p_c|\,,\\
\label{constraint-C-KS}
C^{(1)}_{{\rm ax}}&= 2A^{}_{{\rm ax}}\alpha\beta +2B^{}_{{\rm ax}}\gamma\delta +C^{}_{{\rm ax}}\left(\alpha\delta +\beta\gamma\right) +2\left( \alpha \delta^{\prime} -\gamma \beta^{\prime} \right)= \,0\,.
\end{align}
Here, we have used again the notation for derivatives introduced in Subsec. \ref{Ss:diagonalizing-canonical}, with the prime symbol, and we have called $V_{\ell}^{{\rm ax}}$ the axial potential up to the factor $|p_c|$, with the term containing the squared Fourier frequency removed. Note that, with respect to the sign convention of Sec.~\ref{Ss:Darboux-covariance}, we choose the interior sign on the potential, but this is only to avoid carrying a double sign in our expressions. In the three lines of the above equation, the right-hand side equalities are the conditions that we impose on the new coefficients to reach a diagonal Hamiltonian with a potential in which the only Fourier contribution is exactly a squared term. Explicitly, the new Hamiltonian is
\begin{equation}
       \kappa H^{(1)}_{{\rm ax}}[\tilde{N}] =  \sum_{\mathfrak{n},\lambda}\frac{\tilde{N}|p_c|}{2}\left[ \left({}^{(1)} P_1^{\mathfrak{n},\lambda}\right)^2 +   \left(\omega_n^2 -V_{\ell}^{{\rm ax}}\right)\left({}^{(1)}\! Q_1^{\mathfrak{n},\lambda}\right)^2 \right] \,.
       \label{H-diagonal-axial}
\end{equation}

The reason for introducing the factor $|p_c|$ in Eqs.~\eqref{constraint-A-KS} and~\eqref{constraint-B-KS} becomes clear when considering the equations of motion for the configuration variable ${}^{(1)}\! Q_1^{\mathfrak{n},\lambda} = \Psi^{\mathfrak{n},\lambda}$. Thanks to that factor, the equation takes the desired waveform that we find for axial master functions, namely, 
\begin{equation}
    \left[\partial_\sigma^2 + \omega_n^2 - V_{\ell}^{{\rm ax}}\right] \Psi^{\mathfrak{n},\lambda} = 0 \,.
    \label{eom-diagonal}
\end{equation}
As we previously defined it (see also Appendix~\ref{App: LB}), $\text{d}\chi = |p_c|\tilde{N}\text{d}\sigma$ is a generalization of the tortoise coordinate which reduces to the standard definition in Schwarzschild case, i.e. $\text{d}\sigma= \text{d}\chi/f$. It is radial in the exterior and timelike in the interior. This change of evolution parameter serves well for the aim of connecting with DTs, but it will be easy to restore the original evolution coordinate in Sec.~\ref{Ss:GS-master}. 

The constraint in Eq.~\eqref{constraint-A-KS} reflects the requirement of a frequency independent potential when the dynamical equation is written as a wave equation determined by a Laplacian for the two-dimensional set of symmetry orbits. This frequency independence is crucial in order to have a clear physical understanding of the transformation among the admissible potentials without introducing a nontrivial nonlocality in the system~\cite{ Glampedakis:2017rar,Yurov:2018ynn} (see also Sec.~\ref{Ss:Darboux-covariance}). It turns out that the allowed Hamiltonians are Darboux related as described in Sec.~\ref{Ss:Darboux-covariance}. The requirement that the only frequency dependent contribution to $A^{(1)}_{{\rm ax}}$ is $|p_c| \omega_n^2$ fixes the canonical transformation coefficient $\alpha$ to
\begin{equation}
\alpha_*^{(\pm)} = -\frac{4\ell(\ell+1)L_o^2 }{p_b^2}\Omega_b\gamma \pm\sqrt{\frac{(\ell+2)!}{(\ell-2)!|p_c|}}
\,.
\label{alphastar}
\end{equation}
Taking this into account we can exploit the general results obtained in Sec.~\ref{Ss:diagonalizing-canonical}, in particular Eq.~\eqref{V-diagonal}, and rewrite the condition on $A^{(1)}_{{\rm ax}}$ in Eq.~\eqref{constraint-A-KS} in the following form:
\begin{equation}
    \omega^2 -V_{\ell}^{{\rm ax}} = \frac{B_{{\rm ax}}}{|p_c| \beta} -w^2 -\partial_\sigma w\,,\quad w= \frac{\alpha_*^{(\pm)}}{\beta}\,.
    \label{VGDT-alpha-fixed}
\end{equation}
Then, from the results of Sec.~\ref{S:darboux-symmetry} it is clear that this describes a family of potentials related to each other by a generalized DT (in the sense defined in Sec.~\ref{Ss:darboux-canonical}), which contains infinite Darboux related potentials with the same trivial constant dependence on the squared Fourier frequency as the relevant subclass of transformations which are interesting for our purposes. 

On the other hand, by simply substituting, e.g. $\alpha_*^{(-)}$ into Eq.~\eqref{constraint-A-KS}, we can write the potential $V_{\ell}^{{\rm ax}}$ in the following form:
\begin{equation}
    V_{\ell}^{{\rm ax}} = V_{\rm RW} - \tilde{\gamma}^2 + \partial_\sigma\tilde{\gamma}	\,,\quad   \gamma=\sqrt{\frac{(\ell-2)!}{(\ell+2)!}|p_c|}\left(\frac{\Omega_b}{|p_c|}-\tilde{\gamma}\right)\,,
    \label{Vax-alpha-fixed}
\end{equation}
where $ V_{\rm RW} $ denotes the Regge-Wheeler potential, which in terms of the background variables used in our description can be expressed as~\cite{new}
\begin{equation}
    V_{\rm RW}=  -\frac{1}{p_c^2}\left[\ell(\ell+1)\frac{p_b^2}{L_o^2} - 3\left(\Omega_b^2+\frac{p_b^2}{L_o^2}\right)\right]\,.
    \label{RW-potential-tetrad}
\end{equation}
In particular, when evaluated for a Schwarzschild spacetime with the use of Eq.~\eqref{schwarzschild-tetrad}, it adopts the well-known expression (with the interior and exterior corresponding to $2M/\chi$ larger or smaller than 1)
\begin{equation}
    V_{\rm RW}=  \left(1-\frac{2M}{\chi}\right)\left(\frac{\ell(\ell+1)}{\chi^2} -\frac{6M}{\chi^3}\right)\,.
    \label{RW-potential}
\end{equation}

It is worth noting that the only requirement~\eqref{constraint-A-KS} results in the potentials~\eqref{Vax-alpha-fixed}, which have an interesting structure. Indeed, we have singled out the Regge-Wheeler potential, plus a superpotential $\tilde{\gamma}$, from the quite general condition~\eqref{constraint-A-KS}. Moreover, it can be recast in the following suggestive form, in terms of a single superpotential $W$:
\begin{equation}
    V_{\ell}^{{\rm ax}} = -\partial_\sigma W + W^2 +\mathbf{c}\,,\quad W =  G -\tilde{\gamma}  \,,
    \label{superpotential-Riccati}
\end{equation}
where $G$ has to satisfy the following Riccati equation:
\begin{equation}
    -\partial_\sigma  G + G^2 +2\tilde{\gamma} \left( G -\tilde{\gamma}\right)  + \mathbf{c} = V_{\rm RW} \,.
    \label{Riccati-G}
\end{equation}
The fact that we need this supplementary equation to write explicitly the potentials in terms of superpotentials as in a DT is not surprising. In fact, we already explained that Eq.~\eqref{VGDT-alpha-fixed} describes a family of potentials related by generalized DTs, so that Eq.~\eqref{Riccati-G} represents the restriction to Darboux related potentials. Indeed, the constraint in Eq.~\eqref{constraint-A-KS} guarantees the frequency independence of the transformation, since $\alpha_*^{(-)}$ is independent, and so is $\gamma$ as a consequence. One can check that the rest of the equations are frequency independent with this choice as well. In particular, Eq.~\eqref{Vax-alpha-fixed} shows that we obtain the same potentials which can be found by DTs starting with a specific one, for example the Regge-Wheeler potential corresponding to $\bar{\gamma}=0$.

At this point, we still have to solve the constraints $B^{(1)}_{{\rm ax}}=|p_c|$ and $C^{(1)}_{{\rm ax}}=0$, given in Eqs.~\eqref{constraint-B-KS} and~\eqref{constraint-C-KS}, together with the condition on the coefficients of the transformation so that it be canonical. Finally, one can find the superpotential by solving the Riccati equation~\eqref{Riccati-G} for $G$, which depends on such coefficients of the canonical transformation. Given the solution $\alpha_*^{(-)}$, we find a special solution for $\beta$, $\gamma$, and $\delta$ from the constraint $C^{(1)}_{{\rm ax}}=0$~\eqref{constraint-C-KS} and the canonical nature of the transformation, namely,
\begin{equation}
\beta_*^{(-)} = \sqrt{\frac{\ell(\ell+1)}{(\ell +2)(\ell-1)} |p_c|}\, \frac{4 L_o^2  }{p_b^2}\Omega_b\,,\quad
\gamma_*^{(-)}= - \sqrt{\frac{(\ell-2)!}{(\ell+2)!|p_c|}}\,\Omega_b\,,\quad \delta_*^{(-)}=  -\sqrt{\frac{(\ell-2)!}{(\ell+2)!}|p_c|}\, .
\label{beta-gamma-delta-star}
\end{equation}
It is easy to check that the constraint for $B^{(1)}_{{\rm ax}}=0$ in Eq.~\eqref{constraint-B-KS} is automatically satisfied. This solution implies that $\tilde{\gamma}=0$, and therefore, thanks to Eq.~\eqref{Vax-alpha-fixed}, $V_{\ell}^{{\rm ax}} = V_{\rm RW}$ as expected. The Riccati equation~\eqref{Riccati-G} becomes
\begin{equation}
-\partial_\sigma  G + G^2 + \mathbf{c}= V_{\rm RW}
\,,
\label{Riccati-W-RW}
\end{equation}
and is solved in the (interior and exterior of the) Schwarzschild case by
\begin{equation}
G_*^{(-)} = i \sqrt{\mathbf{c}} + \frac{6\,M f(\chi)}{\lambda(\chi)\chi^2} \,,\quad  i \sqrt{\mathbf{c}} = \frac{1}{12 M}\frac{(\ell+2)!}{(\ell-2)!}\,,
\label{superpotential-standard}
\end{equation}
which consistently coincides with the known superpotential for the Regge-Wheeler potential in Schwarzschild case, obtained from the so-called algebraically special solution~\cite{Glampedakis:2017rar, Lenzi:2021njy}.

In conclusion, we followed a Hamiltonian procedure to obtain the Hamiltonians for the mode variables of master functions in spherically symmetric spacetimes by requiring: i) a gauge invariant configuration variable; ii) a diagonal Hamiltonian; iii) a constant coefficient in front of the squared momentum (up to a redefinition of the evolution parameter). We showed that this produces an infinite family of Hamiltonians, related by generalized DTs. Furthermore, after imposing a further condition, namely that, with the choice of evolution parameter arising from the third condition above, the Fourier contribution to the Hamiltonian be constant for each mode (and given by the squared frequency), we obtained a restricted family of transformations which admit a clear physical interpretation and that are in one-to-one correspondence with the infinite family of canonically DTs. For this family, the corresponding Hamiltonians lead to equations of motion that reproduce the master equations of Refs.~\cite{Lenzi:2021wpc,Lenzi:2021njy,Lenzi:2024tgk}. 

\subsection{Regge-Wheeler master equation}
\label{Ss:RW-equation}

In the previous Sections we showed that, once we managed to identify the phase space sector of perturbative gauge invariants, the canonical transformations leading to a diagonal Hamiltonian for these perturbations according to the constraints~\eqref{constraint-A-KS},~\eqref{constraint-B-KS}, and~\eqref{constraint-C-KS} produce a family of Darboux related Hamiltonians. This is clear from Eq.~\eqref{superpotential-Riccati}. Among these Hamiltonians, the one for which the equation of motion is the Regge-Wheeler equation is a specific case. While we already obtained the Regge-Wheeler potential (see also Ref. \cite{new}), it is still necessary to show that the configuration mode variables $\Psi^{\mathfrak{n},\lambda}$ actually are the Fourier coefficients of the Regge-Wheeler master function (for given harmonic labels), when this function is expressed explicitly in terms of the axial perturbations of the metric and their first order derivatives.

We begin with the gauge invariant quantity satisfying the Regge-Wheeler equation, as defined by the canonical transformation with coefficients $\alpha_*^{(-)}$, $\beta_*^{(-)}$, $\gamma_*^{(-)}$, and $\delta_*^{(-)}$, and the corresponding potential given by Eq.~\eqref{RW-potential}. We then follow the Hamiltonian analysis and all the canonical transformations performed up to this point in order to express the analyzed gauge invariant in terms of the original metric perturbations and their first order derivatives. A straightforward computation yields the following expression for the configuration mode variables:
\begin{equation}
    \label{canonical-undo}
   \Psi^{\mathfrak{n},\lambda}={}^{(1)}\! Q_1^{\mathfrak{n},\lambda} = \sqrt{\frac{(\ell-2)!}{(\ell+2)!}}\sqrt{|p_c|}\left[p_1^{\mathfrak{n},\lambda} - 4\ell(\ell+1)\frac{L_o^2}{p_b^2}\Omega_b q_1^{\mathfrak{n},\lambda}\right].
\end{equation}
This expression already represents a gauge invariant quantity in terms of the original perturbative variables. However, to match it with the master function used in BH perturbation theory (and specifically when the background is given by the Schwarzschild solution), further work is required. In particular, we need to express the perturbative momentum in terms of the metric perturbations and their derivatives. For this purpose, the Hamiltonian and perturbative constraints provided in Appendix~\ref{App: iH} are essential. In fact, since we are dealing with equations of motion, we are no longer working off shell. To be more precise, we are no longer off shell with respect to the perturbations. However, the background is still treated off shell. This means in particular that we are free to set any perturbative constraint to zero and make use of the corresponding on-shell relations. As a first step toward deriving the Regge-Wheeler master function, we employ the perturbative constraint~\eqref{constraint-C0} to obtain the expression
\begin{equation}
    \lambda\omega_n{}^{(1)}\! Q_1^{\mathfrak{n},-\lambda} = -\sqrt{\frac{(\ell-2)!}{(\ell+2)!}}\sqrt{|p_c|}\left[2 p_2^{\mathfrak{n},\lambda} - \frac{(\ell+2)!}{(\ell-2)!}(\Omega_b + \Omega_c)\frac{q_2^{\mathfrak{n},\lambda}}{p_c^2}\right].
    \label{lambdaQ1}
\end{equation}
We further proceed by using the following equation of motion:
\begin{equation}
    \frac{1}{2}\frac{\delta}{\delta p_2^{\mathfrak{n},\lambda}} \left[{}^{\rm (ax)}\Delta^2_1[S]\right] = \text{d}q_2^{\mathfrak{n},\lambda} - \left[2\tilde{N}\left(2\frac{(\ell-2)!}{(\ell+2)!}p_c^2 p_2^{\mathfrak{n},\lambda} -\Omega_c q_2^{\mathfrak{n},\lambda}\right) - 2\kappa h_0^{\mathfrak{n},\lambda}\right]\text{d}\chi= 0 .
\end{equation}
When deriving the equations of motion by taking variations of the action, it becomes clear that the constraint also plays an active role in the dynamical description. This important fact may go unnoticed if the Hamiltonian formalism is treated carelessly, especially when the constraints are not properly accounted for. After substituting into Eq.~\eqref{lambdaQ1} the perturbative momentum $p_2^{\mathfrak{n},\lambda}$, obtained from the above constraint, we obtain
\begin{equation}
    \lambda\omega_n{}^{(1)}\! Q_1^{\mathfrak{n},-\lambda}= \sqrt{\frac{(\ell+2)!}{(\ell-2)!}}\frac{1}{\tilde{N}|p_c|^{3/2}}\left[-\kappa h_0^{\mathfrak{n},\lambda} - \frac{1}{2}\partial_\chi q_2^{\mathfrak{n},\lambda} + \tilde{N}\Omega_b q_2^{\mathfrak{n},\lambda}\right].
\end{equation}
In order to finally connect with the Regge-Wheeler master function, let us sum over all Fourier modes, noting that the left-hand side is a derivative in the $\zeta$ direction in which we have expanded in a Fourier series our perturbations [see Eq.~\eqref{Fourier-inversion}]. We then find that
\begin{equation}
  \sqrt{\frac{(\ell-2)!}{(\ell+2)!}}  \partial_\zeta {}^{(1)}\! Q_1^{\ell m} = \frac{1}{\tilde{N}|p_c|^{3/2}}\left[ -\kappa h_0^{\ell m} - \frac{1}{2}\partial_\chi q_2^{\ell m} + \Omega_b\tilde{N}q_2^{\ell m}\right]= \Psi_{\text{RW}}^{\ell m} \,.
\end{equation}
Hence, this is in fact the Regge-Wheeler master function generalized to any background spacetime of the type~\eqref{KS-metric}. A more explicit form connecting with the known expression is obtained in the case of a Schwarzschild background, using Eqs.~\eqref{schwarzschild-tetrad} and~\eqref{variables-redef},
\begin{equation}
    \Psi_{\text{RW}}^{\ell m}\big|_{\text{Sch}} = \frac{f}{\chi}\left[ -\kappa h_0^{\ell m} - \frac{1}{2}\partial_\chi q_2^{\ell m} + \frac{1}{\chi}q_2^{\ell m}\right]=  \frac{f}{\chi}\left[ h_\chi^{\ell m} - \frac{1}{2}\partial_\chi h_2^{\ell m} + \frac{1}{\chi}h_2^{\ell m}\right]\,,
\end{equation}
where the subscript $\text{Sch}$ stands for evaluation on the Schwarzschild solution. Then, the only difference between the Regge-Wheeler master function for the interior and exterior Schwarzschild geometries is the range of $f(\chi)$, which changes from negative to positive, and the spacetime character of the $\chi$ coordinate, either timelike or spacelike.

It is interesting to show that one can obtain the CPM master function in a similar way. Consider the following equation of motion:
\begin{equation}
    \frac{1}{2}\frac{\delta}{\delta p_1^{\mathfrak{n},\lambda}} \left[{}^{\rm (ax)}\Delta^2_1[S]\right] = \text{d}q_1^{\mathfrak{n},\lambda} - \left[\kappa\lambda\omega_n h_0^{\mathfrak{n},-\lambda} - \tilde{N}\left(2\Omega_b q_1^{\mathfrak{n},\lambda}-\frac{1}{\ell(\ell+1)}\frac{p_b^2}{L_o^2}p_1^{\mathfrak{n},\lambda}\right)\right]\text{d}\chi= 0. 
\end{equation}
Repeating the same procedure and substituting the above on-shell identity into Eq. \eqref{canonical-undo}, we get 
\begin{equation}
    {}^{(1)}\! Q_1^{\mathfrak{n},\lambda} =\ell(\ell+1)\sqrt{\frac{(\ell-2)!}{(\ell+2)!} |p_c|}\frac{L_o^2}{p_b^2 \tilde{N}}\left[\partial_{\chi} q_1^{\mathfrak{n},\lambda} -\kappa\lambda\omega_n h_0^{\mathfrak{n},-\lambda} - 2\tilde{N}\Omega_b q_1^{\mathfrak{n},\lambda}\right] =   \frac{1}{2}\sqrt{\frac{(\ell+2)!}{(\ell-2)!}}\Psi_{\rm CPM}^{\mathfrak{n},\lambda} .
\end{equation}
Indeed, after summing over the Fourier modes and performing some minor redefinitions, we obtain a generalized version of the CPM master function. This quantity can be expressed in terms of either the initial or the final axial perturbative coefficients as
\begin{equation}
     2\sqrt{\frac{(\ell-2)!}{(\ell+2)!}}{}^{(1)}\! Q_1^{\ell m} = \frac{2\sqrt{|p_c|}}{(\ell+2)(\ell-1)}\frac{L_o^2}{p_b^2 \tilde{N}}\left[\partial_\chi q_1^{\ell m} - \kappa \partial_\zeta h_0^{\ell m} - 2\tilde{N}\Omega_b q_1^{\ell m}\right] = \Psi_{\text{CPM}}^{\ell m} \,.
\end{equation}
As before, to explicitly confirm that the above expression corresponds to the CPM master function when the Schwarzschild solution is imposed, let us substitute the corresponding background expressions~\eqref{schwarzschild-tetrad} and the rescaling in Eq.~\eqref{variables-redef},
\begin{equation}
    \Psi_{\text{CPM}}^{\ell m}\big|_{\text{Sch}} = \frac{2\chi}{(\ell+2)(\ell-1)}\left[\partial_\chi h_\zeta^{\ell m} - \partial_\zeta h_\chi^{\ell m} - \frac{2}{\chi} h_\zeta^{\ell m}\right]\,.
\end{equation}
Again, we attain a unified description of the master function both for the exterior and interior spacetimes.

To conclude, let us make some final remarks regarding the relation between the two discussed invariants, namely
\begin{equation}
    \Psi_{\text{RW}}^{\ell m} = \frac{1}{2}\partial_\zeta\Psi_{\text{CPM}}^{\ell m}\,,
\end{equation}
Hence, in the absence of source terms (as in our analysis), the two invariants are related by a derivative in the direction of the additional Killing vector field of the background. When perturbative sources are present, the relation between the two master functions is modified by the addition of terms containing only the harmonic components of the energy-momentum tensor. The explicit generalization can be found in Ref.~\cite{Lenzi:2024tgk}. Moreover, within the context of our Hamiltonian formulation, it is straightforward to verify that both master functions satisfy the same type of generalized Regge-Wheeler equation,
\begin{equation}
    \left[\square^{(2)} -\Theta_{\ell}^{{\rm ax}}\right]\Psi_{\text{RW}}^{\ell m} = 0\,, \qquad \left[\square^{(2)} -\Theta_{\ell}^{{\rm ax}}\right]\Psi_{\text{CPM}}^{\ell m} = 0\,.
\end{equation}
Here, $\Theta_{\ell}^{{\rm ax}}$ is related to $V_{\ell}^{{\rm ax}}$ as in Eq.~\eqref{V-Theta}. In fact, the two master functions are related to ${}^{(1)}\! Q_1^{\ell m}$ by a constant factor (CPM) or a derivative in the Killing, orthogonal direction to the evolution (Regge-Wheeler), neither of which affects the solutions for the equations of motion. Recall that the Laplace-Beltrami operator in two dimensions is defined in Appendix~\ref{App: LB}.

\subsection{Metric reconstruction through the inverse canonical transformation}
\label{Ss:metric-recon}

Now that we have determined the expressions for the mode variables of the axial master function, we can compose the two introduced canonical transformations (one to reach gauge invariant variables and another to pass to the considered master function) and obtain the expressions for the metric perturbations in terms of such master variables. To begin, let us write explicitly the first canonical transformation (the one which was implicit in Sec.~\ref{Ss:perturbative-GI}),
\begin{equation}
    \label{eq: recons}
    \begin{aligned}
        &q_1^{\mathfrak{n,\lambda}} = P_1^{\mathfrak{n,\lambda}} - \frac{\lambda\omega_n}{2}Q_2^{\mathfrak{n,-\lambda}},\\
        &q_2^{\mathfrak{n,\lambda}} = Q_2^{\mathfrak{n,\lambda}},\\
        &p_1^{\mathfrak{n,\lambda}} = -Q_1^{\mathfrak{n,\lambda}} - 2\lambda\omega_n \ell(\ell+1)\frac{L_o^2}{p_b^2}\Omega_b Q_2^{\mathfrak{n,-\lambda}},\\
        &p_2^{\mathfrak{n,\lambda}} = P_2^{\mathfrak{n,\lambda}} + \frac{\lambda\omega_n}{2}Q_1^{\mathfrak{n,-\lambda}} + 2\lambda\omega_n \ell(\ell+1)\frac{L_o^2}{p_b^2}\Omega_bP_1^{\mathfrak{n,-\lambda}} + \frac{1}{2}\frac{(\ell+2)!}{(\ell-2)!p_c^2}(\Omega_b+\Omega_c)Q_2^{\mathfrak{n,\lambda}},\\
        &\tilde{h}_0^{\mathfrak{n},\lambda} = h_0^{\mathfrak{n},\lambda} - \frac{\tilde{N}}{\kappa}\frac{(\ell-2)!}{(\ell+2)!}p_c^2\left[P_2^{\mathfrak{n},\lambda} + \lambda\omega_n Q_1^{\mathfrak{n},-\lambda} + 4\lambda\omega_n \ell(\ell+1)\frac{L_o^2}{p_b^2}\Omega_bP_1^{\mathfrak{n},-\lambda} + \frac{(\ell+2)!}{(\ell-2)!p_c^2}\Omega_bQ_2^{\mathfrak{n},\lambda}\right].
    \end{aligned}
\end{equation}
We have included here, for completeness, the briefly mentioned redefinition of the perturbative Lagrange multipliers used in our formalism. Taking these relations into account, along with the canonical transformation used to introduce the Regge-Wheeler master function (see Sec. \ref{Ss:KS-master}), we can rewrite the original perturbative coefficients in Eq. \eqref{eq: recons} in terms of the gauge-invariant quantities proportional to the axial master function. This leads to the following reconstructed expression: 
\begin{equation}
    \begin{aligned}
        &q_1^{\mathfrak{n,\lambda}} = -\sqrt{\frac{(\ell-2)!}{(\ell+2)!}}\left(\sqrt{|p_c|}{}^{(1)}\! P_1^{\mathfrak{n,\lambda}} + \frac{\Omega_b}{\sqrt{|p_c|}}{}^{(1)}\! Q_1^{\mathfrak{n,\lambda}}\right) - \frac{\lambda\omega_n}{2} Q_2^{\mathfrak{n,-\lambda}},\\
        &q_2^{\mathfrak{n,\lambda}} = Q_2^{\mathfrak{n,\lambda}},\\
        &p_1^{\mathfrak{n,\lambda}} = -\sqrt{\frac{\ell(\ell+1)}{|p_c|}}\left((\ell+2)(\ell-1)-4\frac{L_o^2}{p_b^2}\Omega_b^2\right){}^{(1)}\! Q_1^{\mathfrak{n,\lambda}} - 4\sqrt{\frac{\ell(\ell+1)}{(\ell+2)(\ell-1)}|p_c|}\frac{L_o^2}{p_b^2}\Omega_b{}^{(1)}\! P_1^{\mathfrak{n,\lambda}} \\
        &- 2\lambda\omega_n \ell(\ell+1)\frac{L_o^2}{p_b^2}\Omega_b Q_2^{\mathfrak{n,-\lambda}},\\
        &p_2^{\mathfrak{n,\lambda}} = P_2^{\mathfrak{n,\lambda}} - \frac{\lambda\omega_n}{2}\sqrt{\frac{(\ell+2)!}{(\ell-2)!|p_c|}}{}^{(1)}\! Q_1^{\mathfrak{n,-\lambda}} + \frac{1}{2}\frac{(\ell+2)!}{(\ell-2)!p_c^2}(\Omega_b+\Omega_c)Q_2^{\mathfrak{n,\lambda}}\,.
    \end{aligned}
\end{equation}
In summary, in order to perform the metric reconstruction we had to invert the canonical transformations we used to obtain the desired master variables, i.e.: i) the canonical transformation described in Sec.~\ref{Ss:perturbative-GI} to obtain gauge invariant perturbative variables (see Ref.~\cite{MenaMarugan:2024qnj} for the details on this transformation); and ii) the canonical transformation defined by coefficients in Eqs.~\eqref{alphastar} and~\eqref{beta-gamma-delta-star}. Alternatively, to express these relations solely in terms of the master functions and their first-order derivatives, one may eliminate the momenta of the two independent axial pairs using Eqs. \eqref{liga} and \eqref{H-diagonal-axial}, together with the perturbative Poisson bracket structure, to obtain their corresponding Hamiltonian equations. In this way, ${}^{(1)} P_1^{\mathfrak{n,\lambda}}$ and $P_2^{\mathfrak{n,\lambda}}$ are expressed in terms of derivatives of the configuration variables (up to terms proportional to the perturbative Lagrange multiplier).

\subsection{Gerlach-Sengupta master equation}
\label{Ss:GS-master}

We conclude this Section by showing the relation (up to nonrelevant constant global factors) between the CPM master function and the so-called GS~\cite{Gerlach:1979rw, Gerlach:1980tx} master function. The two are actually related by the following canonical transformation $\{\Psi^{\mathfrak{n},\lambda}, {}^{(1)} P_1^{\mathfrak{n},\lambda} \} \rightarrow \{ \Psi_{\rm GS}^{\mathfrak{n},\lambda} ,{}^{(2)}\! P_1^{\mathfrak{n},\lambda}\}$: 
\begin{equation}
\Psi^{\mathfrak{n},\lambda} = \sqrt{|p_c|}\Psi_{\rm GS}^{\mathfrak{n},\lambda}  \,,\quad
{}^{(1)} P_1^{\mathfrak{n},\lambda} = \frac{1}{\sqrt{|p_c|}}\left( {}^{(2)}\!P_1^{\mathfrak{n},\lambda} + \frac{1}{2}\frac{|p_c|^{\prime}}{|p_c|} \Psi_{\rm GS}^{\mathfrak{n},\lambda} \right)  \, .
\label{canonical-GS}
\end{equation}
Hence, the transformation scales the configuration (master) function by a background function and corrects the canonical momentum by the inverse scaling and the addition of a linear term in the configuration that preserves the diagonal form of the Hamiltonian. We have used again the notation with the prime symbol introduced in Subsec. \ref{Ss:diagonalizing-canonical} for the $\chi$ derivative divided by the densitized lapse. Recall that the derivative in the $\chi$ direction of any function on the background phase space, for example $|p_c|$, can be computed taking Poisson brackets with the background Hamiltonian (see footnote \ref{backdyna}). In this sense, the prime symbol can be understood as a compact notation to simplify our formulas. On the other hand, the scaling introduced above changes the natural parameter in which the Hamiltonian generates the evolution, as we will see. This change of evolution parameter implies a specific modification of the Hamiltonian, that we analyze below.
 
A type-3 generating function for the introduced transformation is
\begin{equation}
F_{(3)}\left({}^{(1)} P_1, \Psi_{\rm GS}\right)= \sum_{\mathfrak{n},\lambda} \left[ -\sqrt{p_c}  \Psi_{\rm GS}^{\mathfrak{n},\lambda} \;{}^{(1)} P_1^{\mathfrak{n},\lambda} +   \frac{1}{4}\frac{|p_c|^{\prime}}{|p_c|} \left(\Psi_{\rm GS}^{\mathfrak{n},\lambda}\right)^2 \right],\, \quad \Psi^{\mathfrak{n},\lambda} = - \frac{\partial F_{(3)}}{\partial {}^{(1)} P_1^{\mathfrak{n},\lambda}}\,,\quad  {}^{(2)} P_1^{\mathfrak{n},\lambda}=- \frac{\partial F_{(3)}}{\partial \Psi_{\rm GS}^{\mathfrak{n},\lambda}}\,.
\end{equation}
Since the transformation depends on the background, the Hamiltonian (apart from the direct effect of the scaling) acquires a form different from that of Eq.~\eqref{H-diagonal-axial}. More precisely, it is modified by a derivative of the generating function (see footnote~\ref{backdyna} for further comments),
\begin{equation}
       \kappa H_{\rm GS}[\tilde{N}] = \sum_{\mathfrak{n},\lambda}\frac{\tilde{N}}{2}\left[ \left({}^{(2)} P_1^{\mathfrak{n},\lambda}\right)^2 +  \left(\omega_n^2p_c^2 + V_{\rm GS}\right)\left(\Psi_{\rm GS}^{\mathfrak{n},\lambda}\right)^2 \right] \,, 
\label{H3-odd}
\end{equation}
where
\begin{equation}
V_{\rm GS} = p_c^2 V_{\rm RW} + \frac{1}{2}\left[ \frac{|p_c|^{\prime\prime}}{|p_c|}-
 \frac{3}{2}\left(\frac{|p_c|^{\prime}}{|p_c|}\right)^2 \right]
\,.
\label{potential-GS}
\end{equation}
It is worth noting that the additional term between square brackets is just the Schwarzian derivative of the function $\sigma(\chi)$ defined by the scaling function $|p_c|$ through the change: $\text{d}\chi = |p_c|\tilde{N}\text{d}\sigma$. In fact, the same type of modification of the Hamiltonian (i.e., a scaling owing to a change in the evolution parameter, and the addition of the corresponding Schwarzian derivative) would also have been obtained if we had scaled the configuration mode variables by a different background function and completed the transformation into a canonical one respecting the diagonal form of the Hamiltonian. More concretely, in our case the transformed potential can be written in the form
\begin{equation}
V_{\rm GS} = \left(\sigma^{\prime}\right)^2 V_{\rm RW} +  \frac{1}{2}\mathcal{S}[\sigma] \,,\quad \mathcal{S}[\sigma]= \frac{\sigma^{\prime\prime\prime}}{\sigma^{\prime}} -\frac{3}{2}\left(\frac{\sigma^{\prime\prime}}{\sigma^{\prime}}\right)^2\,,
\label{V-schwarzian}
\end{equation}
where $\mathcal{S}[\sigma]$ denotes the Schwarzian derivative, a structure that is known to be tightly related to the Korteweg-de Vries equation/Virasoro algebra~\cite{DiFrancesco:1997nk} (see Refs.~\cite{Jaramillo:2024qjz, BenAchour:2022uqo} for the appearance of these structures in BH perturbation theory).

In summary: A) the canonical transformation in Eq.~\eqref{canonical-GS} introduces a scale transformation on the configuration mode variables/master function, thus inducing a change in the potential according to Eq.~\eqref{potential-GS}; B) if we relate the transformation of the potential in Eq.~\eqref{potential-GS} to a transformation of the evolution parameter $\chi \rightarrow \sigma=\sigma(\chi)$, then the potential is scaled by a multiplicative factor and is shifted by the Schwarzian derivative of the transformation. This is a consequence of the fact that the considered canonical transformation simply realizes a conformal transformation in one dimension (the evolution direction), for which the underlying algebraic structure is the Virasoro algebra~\cite{DiFrancesco:1997nk,Jaramillo:2024qjz}. 

In the tetrad representation, the Schwarzian contribution has a very simple form. Indeed, if we use the background Hamiltonian to compute the derivatives with respect to $\chi$ of our background functions (see footnote \ref{backdyna}), we obtain that $|p_c|^{\prime}= 2|p_c|\Omega_b$ and $\Omega_b^{\prime} = -p_b^2/L_o^2$. The new potential can then be written as
\begin{equation}
V_{\rm GS} =\ell(\ell+1)\frac{p_b^2}{L_o^2}  - 4\left(\Omega_b^2+\frac{p_b^2}{L_o^2}\right)\,.
\end{equation}
In fact, it is now easy to see that
\begin{equation}
\mathcal{S} = -2\left(\Omega_b^2+\frac{p_b^2}{L_o^2}\right)
\,.
\label{schwarzian-ashtekar}
\end{equation}
In the specific case of the Schwarzschild spacetime, using Eq.~\eqref{schwarzschild-tetrad}, this Schwarzian term reads
\begin{equation}
\mathcal{S}[\sigma(\chi)] = 2\chi f(\chi)\left[1+f(\chi)\right]
\,.
\end{equation}
It is clear from the expression of the metric~\eqref{KS-metric} that $p_c$ acts as a conformal factor for the orbits of spherical symmetry, while $p_b$ does so for the Lorentzian manifold describing the set of these orbits. It is thus curious to note that the Schwarzian of the conformal transformation determined by $|p_c|$ is actually defined (on shell) by $p_b$ alone [see Eq.~\eqref{schwarzian-ashtekar}].

\section{Conclusion}
\label{conclusion}

In this work we have developed a Hamiltonian formulation for the axial perturbations of a spherically symmetric spacetime in tetrad-connection variables. The use of these variables in this context allows us to provide a unified description of the exterior Schwarzschild spacetime and the interior case. Indeed, while the interior is usually obtained by a Wick rotation of the coordinates, within this perspective exterior and interior are related by a complex canonical transformation on the phase space tetrad-connection variables. This guarantees that all the results we obtain are valid for both regions of the Schwarzschild spacetime, the only difference being in the spacetime character of the evolution parameter for the Hamiltonian flows, which is timelike in the interior but radial in the exterior. As already mentioned, radial Hamiltonians have also been used recently in studies of static symmetries and dynamical features of spherically symmetric spacetimes~\cite{BenAchour:2022uqo,BenAchour:2023dgj,Perez:2023jrq,Livine:2025soz}.

Within this perturbative Hamiltonian framework, we have studied how to reduce the action to correspond to axial gauge invariants satisfying a wave equation, i.e. the so called axial master functions and equations in BH perturbation theory. The Hamiltonian perspective provides a systematic way to obtain them by performing a series of canonical transformations with some general criteria. In particular, the targeted Hamiltonians are required to satisfy three main requirements, in order to describe the dynamics of master functions in terms of wave equations. The first one is that, using a Fourier expansion, one of the canonical pairs of mode variables has to be gauge invariant. This becomes a mechanical task in the Hamiltonian setting since it follows by performing a canonical transformation in which one of the new phase space variables is proportional to the perturbative gauge constraint. Since the constraint generates gauge transformations on first order perturbations and the new canonical pair commutes with it by construction, a gauge invariant pair is selected (see Sec.~\ref{Ss:perturbative-GI}). After this choice has been made, we end up in general with Hamiltonians of the form displayed in Eq.~\eqref{Hamiltonian-QP}. The second and third requirements are that the Hamiltonian be diagonal and that the coefficient in front of the squared momentum term be constant (when a proper evolution parameter is chosen). As we discussed in Sec.~\ref{S:darboux-symmetry}, these conditions restrict the choice of momentum and the freedom to scale the configuration mode variables by a background function. This whole procedure, based on tetrad-connection variables and successive canonical transformations, has been used with a dual purpose. First, we have managed to reformulate the Hamiltonian methods of Ref.~\cite{MenaMarugan:2024qnj} in a way that makes clear contact with the reduction to the axial gauge invariant degrees of freedom, described by master functions, typically considered in BH perturbations. Moreover, the procedure is general enough to also account for the Darboux covariance symmetry~\cite{Lenzi:2021njy} among the infinite master equations for axial BH perturbations~\cite{Lenzi:2021wpc,Lenzi:2024tgk} in terms of canonical transformations between the selected Hamiltonians.

Notice that the splitting into three steps is artificial but necessary to follow the reasoning and obtain the desired results without having to rely on educated guesses which are already based on the knowledge of the final outcome. The splitting makes the procedure systematic and easier to follow. However, once this is done, one should not attribute fundamental properties to any of the individual steps, and the result should be interpreted in its entirety. In other words, we could rephrase the three steps in one by saying that starting from the ADM Hamiltonian for the axial gravitational perturbations, we can perform a canonical transformation that maps the starting Hamiltonian into a diagonal one, for an axial gauge invariant scalar, with constant coefficients in front of the corresponding momentum term (and with a suitable choice of evolution parameter). This type of Hamiltonians describe the dynamics of the axial gravitational perturbative degrees of freedom in terms of a master equation. Moreover, we have shown that such canonical transformation is not unique, but it actually connects an infinite family of Hamiltonians. Restricting the considerations to the case of potentials independent of the Fourier frequencies, in order to avoid nonlocal canonical transformations, the Hamiltonians in this family are actually related by DTs, which now include a specific transformation on the momentum mode variables as well. In this sense, this establishes a one-to-one correspondence between canonical transformations and DTs, and provides a Hamiltonian interpretation of the Darboux covariance hidden symmetry~\cite{Lenzi:2021njy} as a canonical symmetry of the system.

In addition, we have shown that a particular member of this family of Hamiltonians leads to the Regge-Wheeler equation. It is interesting to note how the Hamiltonian formalism seems to select the CPM master function over the Regge-Wheeler one in a more natural way. Indeed, while the CPM invariant is obtained by just expressing the configuration mode variables in terms of the metric perturbations following the canonical transformations, the recovery of the Regge-Wheeler invariant requires the use of the global Hamiltonian constraint. Moreover, by inverting the canonical transformations one has a simple and systematic way to determine the metric reconstruction from the master functions (see Sec.~\ref{Ss:metric-recon}). On the other hand, we have also seen how to obtain the GS master function and how this is related to the CPM invariant by a scale transformation, producing a Schwarzian structure in the potential.

Finally, notice that the results of this work, i.e. the master functions and potentials, have been obtained without the need to solve Hamilton equations for the background. This means that, even if in some cases we have shown explicit results for the Schwarzschild solution, the expressions that we have obtained are valid on any generic spherically symmetric effective background, that may depart from the solutions to GR by the inclusion of effective corrections, for example of quantum nature. Since the dynamics is with respect to a direction that is timelike in the BH interior but radial in the exterior, we see that we can describe backgrounds with modified time evolution in the first case and with modified radial dependence in the second case. It would be interesting to study whether such effective background descriptions, or an extension of our Hamiltonian formalism, allows for applications, for example, to scenarios with BH backgrounds with beyond-GR corrections that can be treated perturbatively~\cite{Endlich:2017tqa,Cardoso:2018ptl,Cano:2019ore, Silva:2024ffz}, or with environmental corrections~\cite{Leung:1999iq, Barausse:2014tra,Cheung:2021bol,Cardoso:2021wlq}. Connected to this generality of the background, a comment has to be made on the physical interpretation of the Darboux symmetry. Indeed, it is known that, in the Schwarzschild background, the Darboux symmetry provides isospectral transformations, meaning that all master functions obtained in this way have the same graybody factors and spectrum of QNMs. However, this statement is not in general true in more general spherically symmetric spacetimes, as isospectrality depends on the asymptotics and details of the potentials (see also discussion on this in Sec.~\ref{Ss:Darboux-covariance}). 

At last, a natural extension of this work is to include the polar sector, which is typically cumbersome because it contains more variables, but should in principle have the same structure, as it appears at the level of master equations. With this in mind, one should also consider the DT among the axial and polar sectors, sometimes called Chandrasekhar's transformation, for completeness. Moreover, our results open a natural path toward canonical quantization schemes for BH perturbations (see e.g. the discussion in Refs. \cite{BGA,MenaMarugan:2024qnj, MenaMarugan:2025anx}). We pursue the investigation of the polar sector and the quantization of the combined system formed by the background and its perturbations in Ref.~\cite{new}.

\acknowledgments

This work was partially supported by MCIN/AEI/10.13019/501100011033 (Spanish Ministry of Science and Innovation) and FSE+ under the Grant No. PID2023-149018NB-C41. A. M.-S. acknowledges support from the PIPF-2023 fellowship from Comunidad Aut\'onoma de Madrid with reference PIPF-2023/TEC-30167. C.F.S. is supported by Contract No. PID2022-137674NB-I00 from MCIN/AEI/10.13039/501100011033 (Spanish Ministry of Science and Innovation) and 2021-SGR-01529 (AGAUR, Generalitat de Catalunya). C.F.S. is also partially supported by the program \textit{Unidad de Excelencia Mar\'{\i}a de Maeztu} (Spanish Ministry of Science and Innovation; Ref.: CEX2020-001058-M).

\appendix

\section{Hamiltonian expressions for axial metric perturbations}
\label{App: iH}

Here we present the lengthy expressions for the perturbative constraints and Hamiltonian that provide the Hamiltonian formulation for the axial metric perturbations which we used as starting point for our analysis in the main text. Our notation is the same that was introduced there. For the axial sector, we have the following single perturbative constraint:
\begin{equation}
    \mathcal{C}_0[h_0^{\mathfrak{n},\lambda}] = \sum_{\mathfrak{n},\lambda} h_0^{\mathfrak{n},-\lambda}\left[ \frac{(\ell+2)!}{(\ell-2)!p_c^2}(\Omega_b + \Omega_c)q_2^{\mathfrak{n},-\lambda} - 2p_2^{\mathfrak{n},-\lambda} + \lambda\omega_n\left( p_1^{\mathfrak{n},\lambda} - 4\ell(\ell+1)\frac{L_o^2}{p_b^2}\Omega_b q_1^{\mathfrak{n},\lambda} \right)\right].
    \label{constraint-C0}
\end{equation}
The corresponding perturbative Lagrange multipliers for each mode contribution to this constraint are  $h_0^{\mathfrak{n},\lambda}$. On the other hand, the axial perturbative Hamiltonian (which gives the quadratic contribution of the axial perturbations to the zero-mode of the total Hamiltonian constraint on the system formed by the background and the perturbations) is 
\begin{equation}
    \begin{aligned}
        \kappa\tilde{H}_{{\rm ax}}[\tilde{N}] &= \sum_{\mathfrak{n},\lambda}\frac{\tilde{N}}{2} \left[\frac{p_b^2}{L_o^2}\frac{[p_1^{\mathfrak{n},\lambda}]^2}{\ell(\ell+1)} + \left(6\Omega_b^2 + 4\Omega_b\Omega_c + \frac{p_b^2}{L_o^2}\ell(\ell+1)\right)\frac{L_o^2}{p_b^2}\ell(\ell+1)[q_1^{\mathfrak{n},\lambda}]^2 - 4\Omega_b q_1^{\mathfrak{n},\lambda}p_1^{\mathfrak{n},\lambda}  - 4\Omega_c q_2^{\mathfrak{n},\lambda}p_2^{\mathfrak{n},\lambda} \right. \\
        & \left.+ 4\frac{(\ell-2)!}{(\ell+2)!}p_c^2[p_2^{\mathfrak{n},\lambda}]^2 + \frac{(\ell+2)!}{4(\ell-2)!p_c^2}\left(2\Omega_b^2 + 4\Omega_c^2 + 4\Omega_b\Omega_c + \omega_n^2 p_c^2\right)[q_2^{\mathfrak{n},\lambda}]^2 + \lambda\omega_n\frac{(\ell+2)!}{(\ell-2)!}q_1^{\mathfrak{n},\lambda}q_2^{\mathfrak{n},-\lambda}\right].
    \end{aligned}
\end{equation}
In comparison with Ref.~\cite{MenaMarugan:2024qnj}, the Hamiltonian expressions differ slightly owing to a different convention for the densitized lapse function. The two lapse functions differ by a factor of $2$, which explains the overall factor of $1/2$ appearing in the Hamiltonian.

\section{Coefficients for the axial perturbative Hamiltonian}
\label{App: HC}

We present here the explicit expressions for the background dependent and mode dependent coefficients of the Hamiltonian given in Eq. \eqref{Hamiltonian-QP}. We have
\begin{equation}
    \begin{aligned}
        &\begin{aligned}
            A_{{\rm ax}} = \frac{1}{\ell(\ell+1)}\frac{p_b^2}{L_o^2}\bigg[1+\frac{\omega_n^2p_c^2}{(\ell+2)(\ell-1)}\frac{L_o^2}{p_b^2}\bigg], 
        \end{aligned}\\
        &\begin{aligned}
            B_{{\rm ax}} = \ell(\ell+1)\frac{L_o^2}{p_b^2}\bigg[8\Omega_b^2 + 8\Omega_b\Omega_c + (\ell^2+\ell+2)\frac{p_b^2}{L_o^2} + \frac{16\omega_n^2p_c^2}{(\ell+2)(\ell-1)}\frac{L_o^2}{p_b^2}\Omega_b^2\bigg],
        \end{aligned}\\
        &\begin{aligned}
            C_{{\rm ax}} = \bigg[4+\frac{8\omega_n^2p_c^2}{(\ell+2)(\ell-1)}\frac{L_o^2}{p_b^2}\bigg]\Omega_b.
        \end{aligned}
    \end{aligned}
\end{equation}
We note that, in order to maintain a consistent density weight throughout the entire Hamiltonian expression, the axial contribution must undergo a type-$1$ canonical transformation of the form $(Q_1^{\mathfrak{n},\lambda},P_1^{\mathfrak{n},\lambda})\mapsto(P_1^{\mathfrak{n},\lambda},-Q_1^{\mathfrak{n},\lambda})$. This transformation was not considered in Ref.~\cite{MenaMarugan:2024qnj}, resulting in a slight rearrangement of the axial coefficients compared to those presented in this work. 

\section{Laplace-Beltrami operators}
\label{App: LB}

We show here the expression of the Laplace-Beltrami operator specialized to the metric~\eqref{KS-metric}. First, we consider the four-dimensional expression given by
\begin{equation}
    \square^{(4)} T = \frac{1}{\sqrt{\text{det}(g^{(4)})}}\partial_{\mu}\left(\sqrt{\text{det}(g^{(4)})}g^{(4)\mu\nu}\partial_{\nu}T\right)= - \frac{L_o^2}{p_b^2}\frac{1}{|p_c|} \left[\frac{\partial_{\chi}^2}{\tilde{N}^2}- \left(p_c^2\partial_\zeta^2 + \frac{p_b^2}{L_o^2}\square_{S^2}\right)\right]T ,
\end{equation}
where $T$ is an arbitrary test function, Greek indices from the middle of the alphabet denote spacetime components, and $\square_{S^2}$ is the Laplace-Beltrami operator on the two-sphere. We have explicitly displayed the $\chi$ derivatives, rather than using the compact notation introduced in Subsec. \ref{Ss:diagonalizing-canonical}.

On the other hand, for the two-dimensional reduced metric
\begin{equation}
    g^{(2)} = \tilde{N}\frac{p_b^2}{L_o^2} \left( -|p_c|\tilde{N} \text{d}\chi^2 + \frac{\text{d}\zeta^2}{|p_c|\tilde{N}} \right),
\end{equation}
the corresponding Laplace-Beltrami operator takes the form
\begin{equation}
    \square^{(2)} T = \frac{1}{\sqrt{\text{det}\left(g^{(2)}\right)}}\partial_{\rho}\left(\sqrt{\text{det}\left(g^{(2)}\right)}g^{(2)\rho\varsigma}\partial_{\varsigma}T\right) = \frac{L_o^2}{p_b^2}|p_c|\left[-\frac{\partial_\chi^2}{|p_c|^2\tilde{N}^2}+\partial_\zeta^2\right]T ,
\end{equation}
where the indices $\rho$ and $\varsigma$ stand here for either $\chi$ or $\zeta$. In the evolution coordinate $\sigma$ defined as $|p_c|\tilde{N}\partial_{\sigma} = \partial_\chi$, the operator behaves as the differential operator of a two-dimensional wave equation, up to multiplication by a conformal factor.


\begin{thebibliography}{59}%
\makeatletter
\providecommand \@ifxundefined [1]{%
 \@ifx{#1\undefined}
}%
\providecommand \@ifnum [1]{%
 \ifnum #1\expandafter \@firstoftwo
 \else \expandafter \@secondoftwo
 \fi
}%
\providecommand \@ifx [1]{%
 \ifx #1\expandafter \@firstoftwo
 \else \expandafter \@secondoftwo
 \fi
}%
\providecommand \natexlab [1]{#1}%
\providecommand \enquote  [1]{``#1''}%
\providecommand \bibnamefont  [1]{#1}%
\providecommand \bibfnamefont [1]{#1}%
\providecommand \citenamefont [1]{#1}%
\providecommand \href@noop [0]{\@secondoftwo}%
\providecommand \href [0]{\begingroup \@sanitize@url \@href}%
\providecommand \@href[1]{\@@startlink{#1}\@@href}%
\providecommand \@@href[1]{\endgroup#1\@@endlink}%
\providecommand \@sanitize@url [0]{\catcode `\\12\catcode `\$12\catcode
  `\&12\catcode `\#12\catcode `\^12\catcode `\_12\catcode `\%12\relax}%
\providecommand \@@startlink[1]{}%
\providecommand \@@endlink[0]{}%
\providecommand \url  [0]{\begingroup\@sanitize@url \@url }%
\providecommand \@url [1]{\endgroup\@href {#1}{\urlprefix }}%
\providecommand \urlprefix  [0]{URL }%
\providecommand \Eprint [0]{\href }%
\providecommand \doibase [0]{https://doi.org/}%
\providecommand \selectlanguage [0]{\@gobble}%
\providecommand \bibinfo  [0]{\@secondoftwo}%
\providecommand \bibfield  [0]{\@secondoftwo}%
\providecommand \translation [1]{[#1]}%
\providecommand \BibitemOpen [0]{}%
\providecommand \bibitemStop [0]{}%
\providecommand \bibitemNoStop [0]{.\EOS\space}%
\providecommand \EOS [0]{\spacefactor3000\relax}%
\providecommand \BibitemShut  [1]{\csname bibitem#1\endcsname}%
\let\auto@bib@innerbib\@empty
\bibitem [{\citenamefont {{Bardeen}}(1980)}]{Bardeen:1980PhRvD..22.1882B}%
  \BibitemOpen
  \bibfield  {author} {\bibinfo {author} {\bibfnamefont {J.~M.}\ \bibnamefont
  {{Bardeen}}},\ }\bibfield  {title} {\bibinfo {title} {{Gauge-invariant
  cosmological perturbations}},\ }\href
  {https://doi.org/10.1103/PhysRevD.22.1882} {\bibfield  {journal} {\bibinfo
  {journal} {Phys. Rev. D}\ }\textbf {\bibinfo {volume} {22}},\ \bibinfo
  {pages} {1882} (\bibinfo {year} {1980})}\BibitemShut {NoStop}%
\bibitem [{\citenamefont {Mukhanov}\ \emph {et~al.}(1992)\citenamefont
  {Mukhanov}, \citenamefont {Feldman},\ and\ \citenamefont
  {Brandenberger}}]{Mukhanov:1990me}%
  \BibitemOpen
  \bibfield  {author} {\bibinfo {author} {\bibfnamefont {V.~F.}\ \bibnamefont
  {Mukhanov}}, \bibinfo {author} {\bibfnamefont {H.~A.}\ \bibnamefont
  {Feldman}},\ and\ \bibinfo {author} {\bibfnamefont {R.~H.}\ \bibnamefont
  {Brandenberger}},\ }\bibfield  {title} {\bibinfo {title} {{Theory of
  cosmological perturbations. Part 1. Classical perturbations. Part 2. Quantum
  theory of perturbations. Part 3. Extensions}},\ }\href
  {https://doi.org/10.1016/0370-1573(92)90044-Z} {\bibfield  {journal}
  {\bibinfo  {journal} {Phys. Rep.}\ }\textbf {\bibinfo {volume} {215}},\
  \bibinfo {pages} {203} (\bibinfo {year} {1992})}\BibitemShut {NoStop}%
\bibitem [{\citenamefont {Aghanim}\ \emph {et~al.}(2020)\citenamefont {Aghanim}
  \emph {et~al.}}]{Planck:2018nkj}%
  \BibitemOpen
  \bibfield  {author} {\bibinfo {author} {\bibfnamefont {N.}~\bibnamefont
  {Aghanim}} \emph {et~al.} (\bibinfo {collaboration} {Planck Collaboration}),\ }\bibfield
  {title} {\bibinfo {title} {{Planck 2018 results. I. Overview and the
  cosmological legacy of Planck}},\ }\href
  {https://doi.org/10.1051/0004-6361/201833880} {\bibfield  {journal} {\bibinfo
   {journal} {Astron. Astrophys.}\ }\textbf {\bibinfo {volume} {641}},\
  \bibinfo {pages} {A1} (\bibinfo {year} {2020})}
  \BibitemShut {NoStop}%
\bibitem [{\citenamefont {{Chandrasekhar}}(1992)}]{Chandrasekhar:1992bo}%
  \BibitemOpen
  \bibfield  {author} {\bibinfo {author} {\bibfnamefont {S.}~\bibnamefont
  {{Chandrasekhar}}},\ }\href@noop {} {\emph {\bibinfo {title} {{The
  Mathematical Theory of Black Holes}}}}\ (\bibinfo  {publisher} {Oxford
  University Press},\ \bibinfo {address} {New York},\ \bibinfo {year}
  {1992})\BibitemShut {NoStop}%
\bibitem [{\citenamefont {Futterman}\ \emph {et~al.}(2012)\citenamefont
  {Futterman}, \citenamefont {Handler},\ and\ \citenamefont
  {Matzner}}]{Futterman:1988ni}%
  \BibitemOpen
  \bibfield  {author} {\bibinfo {author} {\bibfnamefont {J.~A.~H.}\
  \bibnamefont {Futterman}}, \bibinfo {author} {\bibfnamefont {F.~A.}\
  \bibnamefont {Handler}},\ and\ \bibinfo {author} {\bibfnamefont {R.~A.}\
  \bibnamefont {Matzner}},\ }\href {https://doi.org/10.1017/CBO9780511735615}
  {\emph {\bibinfo {title} {{Scattering from Black Holes}}}}\ (\bibinfo
  {publisher} {Cambridge University Press, Cambridge, England},\ \bibinfo {year}
  {2012})\BibitemShut {NoStop}%
\bibitem [{\citenamefont {Hawking}(1975)}]{Hawking:1975vcx}%
  \BibitemOpen
  \bibfield  {author} {\bibinfo {author} {\bibfnamefont {S.~W.}\ \bibnamefont
  {Hawking}},\ }\bibfield  {title} {\bibinfo {title} {{Particle creation by black holes}},\ }\href {https://doi.org/10.1007/BF02345020} {\bibfield
  {journal} {\bibinfo  {journal} {Commun. Math. Phys.}\ }\textbf {\bibinfo
  {volume} {43}},\ \bibinfo {pages} {199} (\bibinfo {year} {1975})};\ \href{https://link.springer.com/article/10.1007/BF01608497}{\bibinfo
  {note} {{\bf 46}, 206(E) (1976)}\BibitemShut {NoStop}}%
\bibitem [{\citenamefont {{Nollert}}(1999)}]{Nollert:1999re}%
  \BibitemOpen
  \bibfield  {author} {\bibinfo {author} {\bibfnamefont {H.-P.}\ \bibnamefont
  {{Nollert}}},\ }\bibfield  {title} {\bibinfo {title} {{Quasinormal modes: the
  characteristic `sound' of black holes and neutron stars}},\ }\href {https://doi.org/10.1088/0264-9381/16/12/201}
  {\bibfield  {journal} {\bibinfo  {journal} {Classical Quantum Gravity}\ }\textbf
  {\bibinfo {volume} {16}},\ \bibinfo {pages} {R159} (\bibinfo {year}
  {1999})}\BibitemShut {NoStop}%
\bibitem [{\citenamefont {Kokkotas}\ and\ \citenamefont
  {Schmidt}(1999)}]{Kokkotas:1999bd}%
  \BibitemOpen
  \bibfield  {author} {\bibinfo {author} {\bibfnamefont {K.~D.}\ \bibnamefont
  {Kokkotas}}\ and\ \bibinfo {author} {\bibfnamefont {B.~G.}\ \bibnamefont
  {Schmidt}},\ }\bibfield  {title} {\bibinfo {title} {{Quasinormal modes of
  stars and black holes}},\ }\href {https://doi.org/10.12942/lrr-1999-2}
  {\bibfield  {journal} {\bibinfo  {journal} {Living Rev. Relativity}\ }\textbf
  {\bibinfo {volume} {2}},\ \bibinfo {pages} {2} (\bibinfo {year} {1999})}
  \BibitemShut {NoStop}%
\bibitem [{\citenamefont {Berti}\ \emph {et~al.}(2009)\citenamefont {Berti},
  \citenamefont {Cardoso},\ and\ \citenamefont {Starinets}}]{Berti:2009kk}%
  \BibitemOpen
  \bibfield  {author} {\bibinfo {author} {\bibfnamefont {E.}~\bibnamefont
  {Berti}}, \bibinfo {author} {\bibfnamefont {V.}~\bibnamefont {Cardoso}},\
  and\ \bibinfo {author} {\bibfnamefont {A.~O.}\ \bibnamefont {Starinets}},\
  }\bibfield  {title} {\bibinfo {title} {{Quasinormal modes of black holes and
  black branes}},\ }\href {https://doi.org/10.1088/0264-9381/26/16/163001}
  {\bibfield  {journal} {\bibinfo  {journal} {Classical Quantum Gravity}\ }\textbf
  {\bibinfo {volume} {26}},\ \bibinfo {pages} {163001} (\bibinfo {year}
  {2009})}
  \BibitemShut {NoStop}%
\bibitem [{\citenamefont {Konoplya}\ and\ \citenamefont
  {Zhidenko}(2011)}]{Konoplya:2011qq}%
  \BibitemOpen
  \bibfield  {author} {\bibinfo {author} {\bibfnamefont {R.}~\bibnamefont
  {Konoplya}}\ and\ \bibinfo {author} {\bibfnamefont {A.}~\bibnamefont
  {Zhidenko}},\ }\bibfield  {title} {\bibinfo {title} {{Quasinormal modes of
  black holes: from astrophysics to string theory}},\ }\href
  {https://doi.org/10.1103/RevModPhys.83.793} {\bibfield  {journal} {\bibinfo
  {journal} {Rev. Mod. Phys.}\ }\textbf {\bibinfo {volume} {83}},\ \bibinfo
  {pages} {793} (\bibinfo {year} {2011})}
  \BibitemShut
  {NoStop}%
\bibitem [{\citenamefont {Damour}\ and\ \citenamefont
  {Nagar}(2009)}]{Damour:2009vw}%
  \BibitemOpen
  \bibfield  {author} {\bibinfo {author} {\bibfnamefont {T.}~\bibnamefont
  {Damour}}\ and\ \bibinfo {author} {\bibfnamefont {A.}~\bibnamefont {Nagar}},\
  }\bibfield  {title} {\bibinfo {title} {{Relativistic tidal properties of
  neutron stars}},\ }\href {https://doi.org/10.1103/PhysRevD.80.084035}
  {\bibfield  {journal} {\bibinfo  {journal} {Phys. Rev. D}\ }\textbf {\bibinfo
  {volume} {80}},\ \bibinfo {pages} {084035} (\bibinfo {year} {2009})}
  \BibitemShut {NoStop}%
\bibitem [{\citenamefont {Binnington}\ and\ \citenamefont
  {Poisson}(2009)}]{Binnington:2009bb}%
  \BibitemOpen
  \bibfield  {author} {\bibinfo {author} {\bibfnamefont {T.}~\bibnamefont
  {Binnington}}\ and\ \bibinfo {author} {\bibfnamefont {E.}~\bibnamefont
  {Poisson}},\ }\bibfield  {title} {\bibinfo {title} {{Relativistic theory of
  tidal Love numbers}},\ }\href {https://doi.org/10.1103/PhysRevD.80.084018}
  {\bibfield  {journal} {\bibinfo  {journal} {Phys. Rev. D}\ }\textbf {\bibinfo
  {volume} {80}},\ \bibinfo {pages} {084018} (\bibinfo {year} {2009})}
  \BibitemShut {NoStop}%
\bibitem [{\citenamefont {Amaro-Seoane}\ \emph {et~al.}(2017)\citenamefont
  {Amaro-Seoane} \emph {et~al.}}]{LISA:2017pwj}%
  \BibitemOpen
  \bibfield  {author} {\bibinfo {author} {\bibfnamefont {P.}~\bibnamefont
  {Amaro-Seoane}} \emph {et~al.} (\bibinfo {collaboration} {LISA Collaboration}),\ }\bibfield
   {title} {\bibinfo {title} {{Laser Interferometer Space Antenna}},\
  }\href@noop {} \ \Eprint
  {https://arxiv.org/abs/1702.00786} {arXiv:1702.00786} 
  \BibitemShut {NoStop}%
\bibitem [{\citenamefont {Arun}\ \emph {et~al.}(2022)\citenamefont {Arun} \emph
  {et~al.}}]{LISA:2022kgy}%
  \BibitemOpen
  \bibfield  {author} {\bibinfo {author} {\bibfnamefont {K.~G.}\ \bibnamefont
  {Arun}} \emph {et~al.} (\bibinfo {collaboration} {LISA Collaboration}),\ }\bibfield
  {title} {\bibinfo {title} {{New horizons for fundamental physics with
  LISA}},\ }\href {https://doi.org/10.1007/s41114-022-00036-9} {\bibfield
  {journal} {\bibinfo  {journal} {Living Rev. Relativity}\ }\textbf {\bibinfo
  {volume} {25}},\ \bibinfo {pages} {4} (\bibinfo {year} {2022})}
  \BibitemShut
  {NoStop}%
\bibitem [{\citenamefont {Abac}\ \emph {et~al.}(2025)\citenamefont {Abac} \emph
  {et~al.}}]{ET:2025xjr}%
  \BibitemOpen
  \bibfield  {author} {\bibinfo {author} {\bibfnamefont {A.}~\bibnamefont
  {Abac}} \emph {et~al.} (\bibinfo {collaboration} {ET Collaboration}),\ }\bibfield  {title}
  {\bibinfo {title} {{The science of the Einstein Telescope}},\ }\href@noop {}
  \ \Eprint {https://arxiv.org/abs/2503.12263}
  {arXiv:2503.12263} 
  \BibitemShut {NoStop}%
\bibitem [{\citenamefont {Regge}\ and\ \citenamefont
  {Wheeler}(1957)}]{Regge:1957td}%
  \BibitemOpen
  \bibfield  {author} {\bibinfo {author} {\bibfnamefont {T.}~\bibnamefont
  {Regge}}\ and\ \bibinfo {author} {\bibfnamefont {J.~A.}\ \bibnamefont
  {Wheeler}},\ }\bibfield  {title} {\bibinfo {title} {{Stability of a
  Schwarzschild singularity}},\ }\href
  {https://doi.org/10.1103/PhysRev.108.1063} {\bibfield  {journal} {\bibinfo
  {journal} {Phys. Rev.}\ }\textbf {\bibinfo {volume} {108}},\ \bibinfo {pages}
  {1063} (\bibinfo {year} {1957})}\BibitemShut {NoStop}%
\bibitem [{\citenamefont {{Cunningham}}\ \emph {et~al.}(1978)\citenamefont
  {{Cunningham}}, \citenamefont {{Price}},\ and\ \citenamefont
  {{Moncrief}}}]{Cunningham:1978cp}%
  \BibitemOpen
  \bibfield  {author} {\bibinfo {author} {\bibfnamefont {C.~T.}\ \bibnamefont
  {{Cunningham}}}, \bibinfo {author} {\bibfnamefont {R.~H.}\ \bibnamefont
  {{Price}}},\ and\ \bibinfo {author} {\bibfnamefont {V.}~\bibnamefont
  {{Moncrief}}},\ }\bibfield  {title} {\bibinfo {title} {{Radiation from
  collapsing relativistic stars. I - Linearized odd-parity radiation}},\ }\href
  {https://doi.org/10.1086/156413} {\bibfield  {journal} {\bibinfo  {journal}
  {Astrophys. J.}\ }\textbf {\bibinfo {volume} {224}},\ \bibinfo {pages} {643}
  (\bibinfo {year} {1978})}\BibitemShut {NoStop}%
\bibitem [{\citenamefont {Zerilli}(1970)}]{Zerilli:1970se}%
  \BibitemOpen
  \bibfield  {author} {\bibinfo {author} {\bibfnamefont {F.~J.}\ \bibnamefont
  {Zerilli}},\ }\bibfield  {title} {\bibinfo {title} {{Effective potential for
  even parity Regge-Wheeler gravitational perturbation equations}},\ }\href
  {https://doi.org/10.1103/PhysRevLett.24.737} {\bibfield  {journal} {\bibinfo
  {journal} {Phys. Rev. Lett.}\ }\textbf {\bibinfo {volume} {24}},\ \bibinfo
  {pages} {737} (\bibinfo {year} {1970})}\BibitemShut {NoStop}%
\bibitem [{\citenamefont {Moncrief}(1974)}]{Moncrief:1974vm}%
  \BibitemOpen
  \bibfield  {author} {\bibinfo {author} {\bibfnamefont {V.}~\bibnamefont
  {Moncrief}},\ }\bibfield  {title} {\bibinfo {title} {{Gravitational
  perturbations of spherically symmetric systems. I. The exterior problem}},\
  }\href {https://doi.org/10.1016/0003-4916(74)90173-0} {\bibfield  {journal}
  {\bibinfo  {journal} {Ann. Phys. (N.Y.)}\ }\textbf {\bibinfo {volume} {88}},\
  \bibinfo {pages} {323} (\bibinfo {year} {1974})}\BibitemShut {NoStop}%
\bibitem [{\citenamefont {Gerlach}\ and\ \citenamefont
  {Sengupta}(1979)}]{Gerlach:1979rw}%
  \BibitemOpen
  \bibfield  {author} {\bibinfo {author} {\bibfnamefont {U.~H.}\ \bibnamefont
  {Gerlach}}\ and\ \bibinfo {author} {\bibfnamefont {U.~K.}\ \bibnamefont
  {Sengupta}},\ }\bibfield  {title} {\bibinfo {title} {{Gauge invariant
  perturbations on most general spherically symmetric space-times}},\ }\href
  {https://doi.org/10.1103/PhysRevD.19.2268} {\bibfield  {journal} {\bibinfo
  {journal} {Phys. Rev. D}\ }\textbf {\bibinfo {volume} {19}},\ \bibinfo
  {pages} {2268} (\bibinfo {year} {1979})}\BibitemShut {NoStop}%
\bibitem [{\citenamefont {Gerlach}\ and\ \citenamefont
  {Sengupta}(1980)}]{Gerlach:1980tx}%
  \BibitemOpen
  \bibfield  {author} {\bibinfo {author} {\bibfnamefont {U.~H.}\ \bibnamefont
  {Gerlach}}\ and\ \bibinfo {author} {\bibfnamefont {U.~K.}\ \bibnamefont
  {Sengupta}},\ }\bibfield  {title} {\bibinfo {title} {{Gauge invariant coupled
  gravitational, acoustical, and electromagnetic modes on most general
  spherical space-times}},\ }\href {https://doi.org/10.1103/PhysRevD.22.1300}
  {\bibfield  {journal} {\bibinfo  {journal} {Phys. Rev. D}\ }\textbf {\bibinfo
  {volume} {22}},\ \bibinfo {pages} {1300} (\bibinfo {year}
  {1980})}\BibitemShut {NoStop}%
\bibitem [{\citenamefont {Lenzi}\ and\ \citenamefont
  {Sopuerta}(2021{\natexlab{a}})}]{Lenzi:2021wpc}%
  \BibitemOpen
  \bibfield  {author} {\bibinfo {author} {\bibfnamefont {M.}~\bibnamefont
  {Lenzi}}\ and\ \bibinfo {author} {\bibfnamefont {C.~F.}\ \bibnamefont
  {Sopuerta}},\ }\bibfield  {title} {\bibinfo {title} {{Master functions and
  equations for perturbations of vacuum spherically symmetric spacetimes}},\
  }\href {https://doi.org/10.1103/PhysRevD.104.084053} {\bibfield  {journal}
  {\bibinfo  {journal} {Phys. Rev. D}\ }\textbf {\bibinfo {volume} {104}},\
  \bibinfo {pages} {084053} (\bibinfo {year} {2021}{\natexlab{a}})}
  \BibitemShut
  {NoStop}%
\bibitem [{\citenamefont {Lenzi}\ and\ \citenamefont
  {Sopuerta}(2024)}]{Lenzi:2024tgk}%
  \BibitemOpen
  \bibfield  {author} {\bibinfo {author} {\bibfnamefont {M.}~\bibnamefont
  {Lenzi}}\ and\ \bibinfo {author} {\bibfnamefont {C.~F.}\ \bibnamefont
  {Sopuerta}},\ }\bibfield  {title} {\bibinfo {title} {{Gauge-independent
  metric reconstruction of perturbations of vacuum spherically-symmetric
  spacetimes}},\ }\href {https://doi.org/10.1103/PhysRevD.109.084030}
  {\bibfield  {journal} {\bibinfo  {journal} {Phys. Rev. D}\ }\textbf {\bibinfo
  {volume} {109}},\ \bibinfo {pages} {084030} (\bibinfo {year} {2024})}
  \BibitemShut {NoStop}%
\bibitem [{\citenamefont {Lenzi}\ and\ \citenamefont
  {Sopuerta}(2021{\natexlab{b}})}]{Lenzi:2021njy}%
  \BibitemOpen
  \bibfield  {author} {\bibinfo {author} {\bibfnamefont {M.}~\bibnamefont
  {Lenzi}}\ and\ \bibinfo {author} {\bibfnamefont {C.~F.}\ \bibnamefont
  {Sopuerta}},\ }\bibfield  {title} {\bibinfo {title} {{Darboux covariance: a
  hidden symmetry of perturbed Schwarzschild black holes}},\ }\href
  {https://doi.org/10.1103/PhysRevD.104.124068} {\bibfield  {journal} {\bibinfo
   {journal} {Phys. Rev. D}\ }\textbf {\bibinfo {volume} {104}},\ \bibinfo
  {pages} {124068} (\bibinfo {year} {2021}{\natexlab{b}})}
  \BibitemShut
  {NoStop}%
\bibitem [{\citenamefont {Brizuela}(2011)}]{BrizuelaPhD}%
  \BibitemOpen
  \bibfield  {author} {\bibinfo {author} {\bibfnamefont {D.}~\bibnamefont
  {Brizuela}},\ } {\bibinfo {title} {High-order perturbation theory of spherical spacetimes with application to vacuum and perfect fluid matter}},\
  \href@noop {} {Ph.D. thesis},\ \bibinfo  {school} {Universidad Aut\'onoma de
  Madrid}, \bibinfo {address} {Madrid}, \bibinfo {year} {2011}\BibitemShut
  {NoStop}%
\bibitem [{\citenamefont {Brizuela}\ and\ \citenamefont
  {Mart\'{\i}n-Garc\'{\i}a}(2009)}]{Brizuela:2008sk}%
  \BibitemOpen
  \bibfield  {author} {\bibinfo {author} {\bibfnamefont {D.}~\bibnamefont
  {Brizuela}}\ and\ \bibinfo {author} {\bibfnamefont {J.~M.}\ \bibnamefont
  {Mart\'{\i}n-Garc\'{\i}a}},\ }\bibfield  {title} {\bibinfo {title} {{Hamiltonian theory
  for the axial perturbations of a dynamical spherical background}},\ }\href
  {https://doi.org/10.1088/0264-9381/26/1/015003} {\bibfield  {journal}
  {\bibinfo  {journal} {Classical Quantum Gravity}\ }\textbf {\bibinfo {volume}
  {26}},\ \bibinfo {pages} {015003} (\bibinfo {year} {2009})}
  \BibitemShut
  {NoStop}%
\bibitem [{\citenamefont {Mena~Marug\'an}\ and\ \citenamefont
  {M\'\i{}nguez-S\'anchez}(2024)}]{MenaMarugan:2024qnj}%
  \BibitemOpen
  \bibfield  {author} {\bibinfo {author} {\bibfnamefont {G.~A.}\ \bibnamefont
  {Mena~Marug\'an}}\ and\ \bibinfo {author} {\bibfnamefont {A.}~\bibnamefont
  {M\'\i{}nguez-S\'anchez}},\ }\bibfield  {title} {\bibinfo {title} {{Axial perturbations in Kantowski-Sachs spacetimes and hybrid quantum cosmology}},\
  }\href {https://doi.org/10.1103/PhysRevD.109.106009} {\bibfield  {journal}
  {\bibinfo  {journal} {Phys. Rev. D}\ }\textbf {\bibinfo {volume}
  {109}},\ \bibinfo {pages} {106009} (\bibinfo {year} {2024})}\BibitemShut
  {NoStop}%
\bibitem [{\citenamefont {Mena~Marug{\'a}n}\ and\ \citenamefont
  {M{\'\i}nguez-S{\'a}nchez}(2025)}]{MenaMarugan:2025anx}%
  \BibitemOpen
  \bibfield  {author} {\bibinfo {author} {\bibfnamefont {G.~A.}\ \bibnamefont
  {Mena~Marug{\'a}n}}\ and\ \bibinfo {author} {\bibfnamefont {A.}~\bibnamefont
  {M{\'\i}nguez-S{\'a}nchez}},\ }\bibfield  {title} {\bibinfo {title} {{Polar
  perturbations in Kantowski-Sachs spacetimes and hybrid quantum~cosmology}},\
  }\href {https://doi.org/10.1103/PhysRevD.111.086024} {\bibfield  {journal}
  {\bibinfo  {journal} {Phys. Rev. D}\ }\textbf {\bibinfo {volume} {111}},\
  \bibinfo {pages} {086024} (\bibinfo {year} {2025})}
  \BibitemShut
  {NoStop}%
\bibitem [{\citenamefont {Arnowitt}\ \emph {et~al.}(2008)\citenamefont
  {Arnowitt}, \citenamefont {Deser},\ and\ \citenamefont
  {Misner}}]{Arnowitt:1962hi}%
  \BibitemOpen
  \bibfield  {author} {\bibinfo {author} {\bibfnamefont {R.~L.}\ \bibnamefont
  {Arnowitt}}, \bibinfo {author} {\bibfnamefont {S.}~\bibnamefont {Deser}},\
  and\ \bibinfo {author} {\bibfnamefont {C.~W.}\ \bibnamefont {Misner}},\
  }\bibfield  {title} {\bibinfo {title} {{The dynamics of general
  relativity}},\ }\href {https://doi.org/10.1007/s10714-008-0661-1} {\bibfield
  {journal} {\bibinfo  {journal} {Gen. Rellativ. Gravit.}\ }\textbf {\bibinfo {volume}
  {40}},\ \bibinfo {pages} {1997} (\bibinfo {year} {2008})}
  \BibitemShut {NoStop}%
\bibitem [{\citenamefont {Ben~Achour}\ \emph {et~al.}(2022)\citenamefont
  {Ben~Achour}, \citenamefont {Livine}, \citenamefont {Mukohyama},\ and\
  \citenamefont {Uzan}}]{BenAchour:2022uqo}%
  \BibitemOpen
  \bibfield  {author} {\bibinfo {author} {\bibfnamefont {J.}~\bibnamefont
  {Ben~Achour}}, \bibinfo {author} {\bibfnamefont {E.~R.}\ \bibnamefont
  {Livine}}, \bibinfo {author} {\bibfnamefont {S.}~\bibnamefont {Mukohyama}},\
  and\ \bibinfo {author} {\bibfnamefont {J.-P.}\ \bibnamefont {Uzan}},\
  }\bibfield  {title} {\bibinfo {title} {{Hidden symmetry of the static
  response of black holes: applications to Love numbers}},\ }\href
  {https://doi.org/10.1007/JHEP07(2022)112} {\bibfield  {journal} {\bibinfo
  {journal} {J. High Energy Phys.}} \bibinfo {volume} {07} (\bibinfo {year} {2022}) \bibinfo {pages}
  {112}}
  \BibitemShut {NoStop}%
\bibitem [{\citenamefont {Ben~Achour}\ \emph {et~al.}(2023)\citenamefont
  {Ben~Achour}, \citenamefont {Livine},\ and\ \citenamefont
  {Oriti}}]{BenAchour:2023dgj}%
  \BibitemOpen
  \bibfield  {author} {\bibinfo {author} {\bibfnamefont {J.}~\bibnamefont
  {Ben~Achour}}, \bibinfo {author} {\bibfnamefont {E.~R.}\ \bibnamefont
  {Livine}},\ and\ \bibinfo {author} {\bibfnamefont {D.}~\bibnamefont
  {Oriti}},\ }\bibfield  {title} {\bibinfo {title} {{Schr{\"o}dinger symmetry
  of Schwarzschild-(A)dS black hole mechanics}},\ }\href
  {https://doi.org/10.1103/PhysRevD.108.104028} {\bibfield  {journal} {\bibinfo
   {journal} {Phys. Rev. D}\ }\textbf {\bibinfo {volume} {108}},\ \bibinfo
  {pages} {104028} (\bibinfo {year} {2023})}
  \BibitemShut
  {NoStop}%
\bibitem [{\citenamefont {Perez}\ \emph {et~al.}(2023)\citenamefont {Perez},
  \citenamefont {Ribisi},\ and\ \citenamefont {Viollet}}]{Perez:2023jrq}%
  \BibitemOpen
  \bibfield  {author} {\bibinfo {author} {\bibfnamefont {A.}~\bibnamefont
  {Perez}}, \bibinfo {author} {\bibfnamefont {S.}~\bibnamefont {Ribisi}},\ and\
  \bibinfo {author} {\bibfnamefont {S.}~\bibnamefont {Viollet}},\ }\bibfield
  {title} {\bibinfo {title} {{Modeling quantum particles falling into a black hole: the deep interior limit}},\ }\href
  {https://doi.org/10.3390/universe9020075} {\bibfield  {journal} {\bibinfo
  {journal} {Universe}\ }\textbf {\bibinfo {volume} {9}},\ \bibinfo {pages}
  {75} (\bibinfo {year} {2023})}
  \BibitemShut {NoStop}%
\bibitem [{\citenamefont {Livine}\ and\ \citenamefont
  {Yokokura}(2025)}]{Livine:2025soz}%
  \BibitemOpen
  \bibfield  {author} {\bibinfo {author} {\bibfnamefont {E.~R.}\ \bibnamefont
  {Livine}}\ and\ \bibinfo {author} {\bibfnamefont {Y.}~\bibnamefont
  {Yokokura}},\ }\bibfield  {title} {\bibinfo {title} {{Effective dynamics of
  spherically symmetric static spacetime}},\ }\href
  {https://doi.org/10.1103/djx3-f3ht} {\bibfield  {journal} {\bibinfo
  {journal} {Phys. Rev. D}\ }\textbf {\bibinfo {volume} {112}},\ \bibinfo
  {pages} {104034} (\bibinfo {year} {2025})}
  \BibitemShut
  {NoStop}%
\bibitem [{\citenamefont {Martel}\ and\ \citenamefont
  {Poisson}(2005)}]{Martel:2005ir}%
  \BibitemOpen
  \bibfield  {author} {\bibinfo {author} {\bibfnamefont {K.}~\bibnamefont
  {Martel}}\ and\ \bibinfo {author} {\bibfnamefont {E.}~\bibnamefont
  {Poisson}},\ }\bibfield  {title} {\bibinfo {title} {{Gravitational
  perturbations of the Schwarzschild spacetime: a practical covariant and
  gauge-invariant formalism}},\ }\href
  {https://doi.org/10.1103/PhysRevD.71.104003} {\bibfield  {journal} {\bibinfo
  {journal} {Phys. Rev. D}\ }\textbf {\bibinfo {volume} {71}},\ \bibinfo
  {pages} {104003} (\bibinfo {year} {2005})}
  \BibitemShut
  {NoStop}%
\bibitem [{\citenamefont {Brizuela}\ \emph {et~al.}(2006)\citenamefont
  {Brizuela}, \citenamefont {Mart\'{\i}n-Garc\'{\i}a},\ and\ \citenamefont
  {Mena~Marug\'an}}]{Brizuela:2006ne}%
  \BibitemOpen
  \bibfield  {author} {\bibinfo {author} {\bibfnamefont {D.}~\bibnamefont
  {Brizuela}}, \bibinfo {author} {\bibfnamefont {J.~M.}\ \bibnamefont
  {Mart\'{\i}n-Garc\'{\i}a}},\ and\ \bibinfo {author} {\bibfnamefont {G.~A.}\
  \bibnamefont {Mena~Marug\'an}},\ }\bibfield  {title} {\bibinfo {title} {{Second
  and higher-order perturbations of a spherical spacetime}},\ }\href
  {https://doi.org/10.1103/PhysRevD.74.044039} {\bibfield  {journal} {\bibinfo
  {journal} {Phys. Rev. D}\ }\textbf {\bibinfo {volume} {74}},\ \bibinfo
  {pages} {044039} (\bibinfo {year} {2006})}
\BibitemShut
  {NoStop}%
\bibitem [{\citenamefont {Brizuela}\ \emph {et~al.}(2007)\citenamefont
  {Brizuela}, \citenamefont {Mart\'{\i}n-Garc\'{\i}a},\ and\ \citenamefont
  {Marug\'an}}]{Brizuela:2007zza}%
  \BibitemOpen
  \bibfield  {author} {\bibinfo {author} {\bibfnamefont {D.}~\bibnamefont
  {Brizuela}}, \bibinfo {author} {\bibfnamefont {J.~M.}\ \bibnamefont
  {Mart\'{\i}n-Garc\'{\i}a}},\ and\ \bibinfo {author} {\bibfnamefont {G.~A.}\
  \bibnamefont {Mena~Marug\'an}},\ }\bibfield  {title} {\bibinfo {title} {{High-order
  gauge-invariant perturbations of a spherical spacetime}},\ }\href
  {https://doi.org/10.1103/PhysRevD.76.024004} {\bibfield  {journal} {\bibinfo
  {journal} {Phys. Rev. D}\ }\textbf {\bibinfo {volume} {76}},\ \bibinfo
  {pages} {024004} (\bibinfo {year} {2007})}
  \BibitemShut
  {NoStop}%
\bibitem [{\citenamefont {Yurov}\ and\ \citenamefont
  {Yurov}(2019)}]{Yurov:2018ynn}%
  \BibitemOpen
  \bibfield  {author} {\bibinfo {author} {\bibfnamefont {A.~V.}\ \bibnamefont
  {Yurov}}\ and\ \bibinfo {author} {\bibfnamefont {V.~A.}\ \bibnamefont
  {Yurov}},\ }\bibfield  {title} {\bibinfo {title} {{A look at the generalized
  Darboux transformations for the quasinormal spectra in Schwarzschild black
  hole perturbation theory: just how general should it be?}},\ }\href
  {https://doi.org/10.1016/j.physleta.2019.05.024} {\bibfield  {journal}
  {\bibinfo  {journal} {Phys. Lett. A}\ }\textbf {\bibinfo {volume} {383}},\
  \bibinfo {pages} {2571} (\bibinfo {year} {2019})}
  \BibitemShut
  {NoStop}%
\bibitem [{\citenamefont {Lenzi}\ and\ \citenamefont
  {Sopuerta}(2023{\natexlab{a}})}]{Lenzi:2022wjv}%
  \BibitemOpen
  \bibfield  {author} {\bibinfo {author} {\bibfnamefont {M.}~\bibnamefont
  {Lenzi}}\ and\ \bibinfo {author} {\bibfnamefont {C.~F.}\ \bibnamefont
  {Sopuerta}},\ }\bibfield  {title} {\bibinfo {title} {{Black hole greybody
  factors from Korteweg\textendash{}de Vries integrals: theory}},\ }\href
  {https://doi.org/10.1103/PhysRevD.107.044010} {\bibfield  {journal} {\bibinfo
   {journal} {Phys. Rev. D}\ }\textbf {\bibinfo {volume} {107}},\ \bibinfo
  {pages} {044010} (\bibinfo {year} {2023}{\natexlab{a}})}
  \BibitemShut
  {NoStop}%
\bibitem [{\citenamefont {Lenzi}\ and\ \citenamefont
  {Sopuerta}(2023{\natexlab{b}})}]{Lenzi:2023inn}%
  \BibitemOpen
  \bibfield  {author} {\bibinfo {author} {\bibfnamefont {M.}~\bibnamefont
  {Lenzi}}\ and\ \bibinfo {author} {\bibfnamefont {C.~F.}\ \bibnamefont
  {Sopuerta}},\ }\bibfield  {title} {\bibinfo {title} {{Black hole greybody
  factors from Korteweg\textendash{}de Vries integrals: computation}},\ }\href
  {https://doi.org/10.1103/PhysRevD.107.084039} {\bibfield  {journal} {\bibinfo
   {journal} {Phys. Rev. D}\ }\textbf {\bibinfo {volume} {107}},\ \bibinfo
  {pages} {084039} (\bibinfo {year} {2023}{\natexlab{b}})}
  \BibitemShut
  {NoStop}%
\bibitem [{\citenamefont {{Chandrasekhar}}(1980)}]{1980RSPSA.369..425C}%
  \BibitemOpen
  \bibfield  {author} {\bibinfo {author} {\bibfnamefont {S.}~\bibnamefont
  {{Chandrasekhar}}},\ }\bibfield  {title} {\bibinfo {title} {{On
  one-dimensional potential barriers having equal reflexion and transmission coefficients}},\ }\href {https://doi.org/10.1098/rspa.1980.0008} {\bibfield
  {journal} {\bibinfo  {journal} {Proc. R. Soc. A}\ }\textbf {\bibinfo
  {volume} {369}},\ \bibinfo {pages} {425} (\bibinfo {year}
  {1980})}\BibitemShut {NoStop}%
\bibitem [{\citenamefont {Glampedakis}\ \emph {et~al.}(2017)\citenamefont
  {Glampedakis}, \citenamefont {Johnson},\ and\ \citenamefont
  {Kennefick}}]{Glampedakis:2017rar}%
  \BibitemOpen
  \bibfield  {author} {\bibinfo {author} {\bibfnamefont {K.}~\bibnamefont
  {Glampedakis}}, \bibinfo {author} {\bibfnamefont {A.~D.}\ \bibnamefont
  {Johnson}},\ and\ \bibinfo {author} {\bibfnamefont {D.}~\bibnamefont
  {Kennefick}},\ }\bibfield  {title} {\bibinfo {title} {{Darboux transformation
  in black hole perturbation theory}},\ }\href
  {https://doi.org/10.1103/PhysRevD.96.024036} {\bibfield  {journal} {\bibinfo
  {journal} {Phys. Rev. D}\ }\textbf {\bibinfo {volume} {96}},\ \bibinfo
  {pages} {024036} (\bibinfo {year} {2017})}
  \BibitemShut
  {NoStop}%
\bibitem [{\citenamefont {Lenzi}\ \emph
  {et~al.}(2025{\natexlab{a}})\citenamefont {Lenzi}, \citenamefont {Agudo},\
  and\ \citenamefont {Sopuerta}}]{Lenzi:2025kqs}%
  \BibitemOpen
  \bibfield  {author} {\bibinfo {author} {\bibfnamefont {M.}~\bibnamefont
  {Lenzi}}, \bibinfo {author} {\bibfnamefont {A.~M.}\ \bibnamefont {Agudo}},\
  and\ \bibinfo {author} {\bibfnamefont {C.~F.}\ \bibnamefont {Sopuerta}},\
  }\bibfield  {title} {\bibinfo {title} {{Korteweg-de Vries integrals for
  modified black hole potentials: instabilities and other questions}},\ }\href
  {https://doi.org/10.1088/1475-7516/2025/09/021} {\bibfield  {journal}
  {\bibinfo  {journal} {J. Cosmol. Astropart. Phys.}} \bibinfo {volume} {09} (\bibinfo {year} {2025}) \bibinfo
  {pages} {021}}
  \BibitemShut {NoStop}%
\bibitem [{\citenamefont {{Darboux}}(1882)}]{1999physics...8003D}%
  \BibitemOpen
  \bibfield  {author} {\bibinfo {author} {\bibfnamefont {G.}~\bibnamefont
  {{Darboux}}},\ }\bibfield  {title} {\bibinfo {title} {{Sur une proposition relative aux équations linéaires}},\ }\href{https://arxiv.org/abs/physics/9908003} {\bibfield  {journal}
  {\bibinfo  {journal} {C.R. Acad. Sci. Paris}\ }\textbf {\bibinfo {volume}
  {94}},\ \bibinfo {pages} {1456} (\bibinfo {year} {1882})}
  \BibitemShut {NoStop}%
\bibitem [{\citenamefont {Chandrasekhar}\ and\ \citenamefont
  {Detweiler}(1976)}]{Chandrasekhar:1976zz}%
  \BibitemOpen
  \bibfield  {author} {\bibinfo {author} {\bibfnamefont {S.}~\bibnamefont
  {Chandrasekhar}}\ and\ \bibinfo {author} {\bibfnamefont {S.~L.}\ \bibnamefont
  {Detweiler}},\ }\bibfield  {title} {\bibinfo {title} {{Equations governing gravitational perturbations of the Kerr black-hole}},\ }\href
  {https://doi.org/10.1098/rspa.1976.0101} {\bibfield  {journal} {\bibinfo
  {journal} {Proc. R. Soc. A}\ }\textbf {\bibinfo {volume} {350}},\
  \bibinfo {pages} {165} (\bibinfo {year} {1976})}\BibitemShut {NoStop}%
\bibitem [{\citenamefont {Sasaki}\ and\ \citenamefont
  {Nakamura}(1982)}]{Sasaki:1981kj}%
  \BibitemOpen
  \bibfield  {author} {\bibinfo {author} {\bibfnamefont {M.}~\bibnamefont
  {Sasaki}}\ and\ \bibinfo {author} {\bibfnamefont {T.}~\bibnamefont
  {Nakamura}},\ }\bibfield  {title} {\bibinfo {title} {{A class of new perturbation equations for the Kerr geometry}},\ }\href
  {https://doi.org/10.1016/0375-9601(82)90507-2} {\bibfield  {journal}
  {\bibinfo  {journal} {Phys. Lett.}\ }\textbf {\bibinfo {volume} {89A}},\
  \bibinfo {pages} {68} (\bibinfo {year} {1982})}\BibitemShut {NoStop}%
\bibitem [{\citenamefont {Shah}\ and\ \citenamefont
  {Whiting}(2016)}]{Shah:2015sva}%
  \BibitemOpen
  \bibfield  {author} {\bibinfo {author} {\bibfnamefont {A.~G.}\ \bibnamefont
  {Shah}}\ and\ \bibinfo {author} {\bibfnamefont {B.~F.}\ \bibnamefont
  {Whiting}},\ }\bibfield  {title} {\bibinfo {title} {{Raising and lowering operators of spin-weighted spheroidal harmonics}},\ }\href
  {https://doi.org/10.1007/s10714-016-2064-z} {\bibfield  {journal} {\bibinfo
  {journal} {Gen. Relativ. Gravit.}\ }\textbf {\bibinfo {volume} {48}},\ \bibinfo
  {pages} {78} (\bibinfo {year} {2016})}
  \BibitemShut
  {NoStop}%
\bibitem [{\citenamefont {Lenzi}\ \emph
  {et~al.}(2025{\natexlab{b}})\citenamefont {Lenzi}, \citenamefont
  {M\'{\i}nguez-S\'anchez},\ and\ \citenamefont {Mena~Marug\'an}}]{new}%
  \BibitemOpen
  \bibfield  {author} {\bibinfo {author} {\bibfnamefont {M.}~\bibnamefont
  {Lenzi}}, \bibinfo {author} {\bibfnamefont
  {G.~A.}~\bibnamefont {Mena~Marug\'an}},\ and\ \bibinfo {author} {\bibfnamefont {A.}~\bibnamefont
  {M\'{\i}nguez-S\'anchez}},\ }\bibfield  {title} {\bibinfo {title}
  {{Master functions and hybrid quantization of perturbed nonrotating black
  hole interiors}},\ }\href@noop {} \ \Eprint {https://arxiv.org/abs/2512.10692}
  {arXiv:2512.10692} 
  \BibitemShut {NoStop}%
\bibitem [{\citenamefont {Di~Francesco}\ \emph {et~al.}(1997)\citenamefont
  {Di~Francesco}, \citenamefont {Mathieu},\ and\ \citenamefont
  {Senechal}}]{DiFrancesco:1997nk}%
  \BibitemOpen
  \bibfield  {author} {\bibinfo {author} {\bibfnamefont {P.}~\bibnamefont
  {Di~Francesco}}, \bibinfo {author} {\bibfnamefont {P.}~\bibnamefont
  {Mathieu}},\ and\ \bibinfo {author} {\bibfnamefont {D.}~\bibnamefont
  {Senechal}},\ }\href {https://doi.org/10.1007/978-1-4612-2256-9} {\emph
  {\bibinfo {title} {{Conformal Field Theory}}}},\ Graduate Texts in
  Contemporary Physics\ (\bibinfo  {publisher} {Springer-Verlag},\ \bibinfo
  {address} {New York},\ \bibinfo {year} {1997})\BibitemShut {NoStop}%
\bibitem [{\citenamefont {Jaramillo}\ \emph {et~al.}(2024)\citenamefont
  {Jaramillo}, \citenamefont {Lenzi},\ and\ \citenamefont
  {Sopuerta}}]{Jaramillo:2024qjz}%
  \BibitemOpen
  \bibfield  {author} {\bibinfo {author} {\bibfnamefont {J.~L.}\ \bibnamefont
  {Jaramillo}}, \bibinfo {author} {\bibfnamefont {M.}~\bibnamefont {Lenzi}},\
  and\ \bibinfo {author} {\bibfnamefont {C.~F.}\ \bibnamefont {Sopuerta}},\
  }\bibfield  {title} {\bibinfo {title} {{Integrability in perturbed black
  holes: background hidden structures}},\ }\href
  {https://doi.org/10.1103/PhysRevD.110.104049} {\bibfield  {journal} {\bibinfo
   {journal} {Phys. Rev. D}\ }\textbf {\bibinfo {volume} {110}},\ \bibinfo
  {pages} {104049} (\bibinfo {year} {2024})}
  \BibitemShut
  {NoStop}%
\bibitem [{\citenamefont {Endlich}\ \emph {et~al.}(2017)\citenamefont
  {Endlich}, \citenamefont {Gorbenko}, \citenamefont {Huang},\ and\
  \citenamefont {Senatore}}]{Endlich:2017tqa}%
  \BibitemOpen
  \bibfield  {author} {\bibinfo {author} {\bibfnamefont {S.}~\bibnamefont
  {Endlich}}, \bibinfo {author} {\bibfnamefont {V.}~\bibnamefont {Gorbenko}},
  \bibinfo {author} {\bibfnamefont {J.}~\bibnamefont {Huang}},\ and\ \bibinfo
  {author} {\bibfnamefont {L.}~\bibnamefont {Senatore}},\ }\bibfield  {title}
  {\bibinfo {title} {{An effective formalism for testing extensions to general
  relativity with gravitational waves}},\ }\href
  {https://doi.org/10.1007/JHEP09(2017)122} {\bibfield  {journal} {\bibinfo
  {journal} {J. High Energy Phys.}} \bibinfo {volume} {09} (\bibinfo {year} {2017}) \bibinfo {pages}
  {122}}
  \BibitemShut {NoStop}%
\bibitem [{\citenamefont {Cardoso}\ \emph {et~al.}(2018)\citenamefont
  {Cardoso}, \citenamefont {Kimura}, \citenamefont {Maselli},\ and\
  \citenamefont {Senatore}}]{Cardoso:2018ptl}%
  \BibitemOpen
  \bibfield  {author} {\bibinfo {author} {\bibfnamefont {V.}~\bibnamefont
  {Cardoso}}, \bibinfo {author} {\bibfnamefont {M.}~\bibnamefont {Kimura}},
  \bibinfo {author} {\bibfnamefont {A.}~\bibnamefont {Maselli}},\ and\ \bibinfo
  {author} {\bibfnamefont {L.}~\bibnamefont {Senatore}},\ }\bibfield  {title}
  {\bibinfo {title} {{Black holes in an effective field theory extension of general relativity}},\ }\href
  {https://doi.org/10.1103/PhysRevLett.121.251105} {\bibfield  {journal}
  {\bibinfo  {journal} {Phys. Rev. Lett.}\ }\textbf {\bibinfo {volume} {121}},\
  \bibinfo {pages} {251105} (\bibinfo {year} {2018})}; \href{https://doi.org/10.1103/PhysRevLett.131.109903}{\bibinfo {note}
  {{\bf 131}, 109903(E) (2023)}}
  \BibitemShut
  {NoStop}%
\bibitem [{\citenamefont {Cano}\ and\ \citenamefont
  {Ruip\'erez}(2019)}]{Cano:2019ore}%
  \BibitemOpen
  \bibfield  {author} {\bibinfo {author} {\bibfnamefont {P.~A.}\ \bibnamefont
  {Cano}}\ and\ \bibinfo {author} {\bibfnamefont {A.}~\bibnamefont
  {Ruip\'erez}},\ }\bibfield  {title} {\bibinfo {title} {{Leading
  higher-derivative corrections to Kerr geometry}},\ }\href
  {https://doi.org/10.1007/JHEP05(2019)189} {\bibfield  {journal} {\bibinfo
  {journal} {J. High Energy Phys.}} \bibinfo {volume} {05} (\bibinfo {year} {2019}) \bibinfo {pages}
  {189}}; \bibinfo {note} \href{https://link.springer.com/article/10.1007/JHEP03(2020)187}{{03 (2020) 187(E)}}
  \BibitemShut
  {NoStop}%
\bibitem [{\citenamefont {Silva}\ \emph {et~al.}(2024)\citenamefont {Silva},
  \citenamefont {Tambalo}, \citenamefont {Glampedakis}, \citenamefont {Yagi},\
  and\ \citenamefont {Steinhoff}}]{Silva:2024ffz}%
  \BibitemOpen
  \bibfield  {author} {\bibinfo {author} {\bibfnamefont {H.~O.}\ \bibnamefont
  {Silva}}, \bibinfo {author} {\bibfnamefont {G.}~\bibnamefont {Tambalo}},
  \bibinfo {author} {\bibfnamefont {K.}~\bibnamefont {Glampedakis}}, \bibinfo
  {author} {\bibfnamefont {K.}~\bibnamefont {Yagi}},\ and\ \bibinfo {author}
  {\bibfnamefont {J.}~\bibnamefont {Steinhoff}},\ }\bibfield  {title} {\bibinfo
  {title} {{Quasinormal modes and their excitation beyond general
  relativity}},\ }\href {https://doi.org/10.1103/PhysRevD.110.024042}
  {\bibfield  {journal} {\bibinfo  {journal} {Phys. Rev. D}\ }\textbf {\bibinfo
  {volume} {110}},\ \bibinfo {pages} {024042} (\bibinfo {year} {2024})}
  \BibitemShut {NoStop}%
\bibitem [{\citenamefont {Leung}\ \emph {et~al.}(1999)\citenamefont {Leung},
  \citenamefont {Liu}, \citenamefont {Suen}, \citenamefont {Tam},\ and\
  \citenamefont {Young}}]{Leung:1999iq}%
  \BibitemOpen
  \bibfield  {author} {\bibinfo {author} {\bibfnamefont {P.~T.}\ \bibnamefont
  {Leung}}, \bibinfo {author} {\bibfnamefont {Y.~T.}\ \bibnamefont {Liu}},
  \bibinfo {author} {\bibfnamefont {W.~M.}\ \bibnamefont {Suen}}, \bibinfo
  {author} {\bibfnamefont {C.~Y.}\ \bibnamefont {Tam}},\ and\ \bibinfo {author}
  {\bibfnamefont {K.}~\bibnamefont {Young}},\ }\bibfield  {title} {\bibinfo
  {title} {{Perturbative approach to the quasinormal modes of dirty black
  holes}},\ }\href {https://doi.org/10.1103/PhysRevD.59.044034} {\bibfield
  {journal} {\bibinfo  {journal} {Phys. Rev. D}\ }\textbf {\bibinfo {volume}
  {59}},\ \bibinfo {pages} {044034} (\bibinfo {year} {1999})}
  \BibitemShut
  {NoStop}%
\bibitem [{\citenamefont {Barausse}\ \emph {et~al.}(2014)\citenamefont
  {Barausse}, \citenamefont {Cardoso},\ and\ \citenamefont
  {Pani}}]{Barausse:2014tra}%
  \BibitemOpen
  \bibfield  {author} {\bibinfo {author} {\bibfnamefont {E.}~\bibnamefont
  {Barausse}}, \bibinfo {author} {\bibfnamefont {V.}~\bibnamefont {Cardoso}},\
  and\ \bibinfo {author} {\bibfnamefont {P.}~\bibnamefont {Pani}},\ }\bibfield
  {title} {\bibinfo {title} {{Can environmental effects spoil precision
  gravitational-wave astrophysics?}},\ }\href
  {https://doi.org/10.1103/PhysRevD.89.104059} {\bibfield  {journal} {\bibinfo
  {journal} {Phys. Rev. D}\ }\textbf {\bibinfo {volume} {89}},\ \bibinfo
  {pages} {104059} (\bibinfo {year} {2014})}
  \BibitemShut
  {NoStop}%
\bibitem [{\citenamefont {Cheung}\ \emph {et~al.}(2022)\citenamefont {Cheung},
  \citenamefont {Destounis}, \citenamefont {Macedo}, \citenamefont {Berti},\
  and\ \citenamefont {Cardoso}}]{Cheung:2021bol}%
  \BibitemOpen
  \bibfield  {author} {\bibinfo {author} {\bibfnamefont {M.~H.-Y.}\
  \bibnamefont {Cheung}}, \bibinfo {author} {\bibfnamefont {K.}~\bibnamefont
  {Destounis}}, \bibinfo {author} {\bibfnamefont {R.~P.}\ \bibnamefont
  {Macedo}}, \bibinfo {author} {\bibfnamefont {E.}~\bibnamefont {Berti}},\ and\
  \bibinfo {author} {\bibfnamefont {V.}~\bibnamefont {Cardoso}},\ }\bibfield
  {title} {\bibinfo {title} {{Destabilizing the fundamental mode of black holes: the elephant and the flea}},\ }\href
  {https://doi.org/10.1103/PhysRevLett.128.111103} {\bibfield  {journal}
  {\bibinfo  {journal} {Phys. Rev. Lett.}\ }\textbf {\bibinfo {volume} {128}},\
  \bibinfo {pages} {111103} (\bibinfo {year} {2022})}
  \BibitemShut
  {NoStop}%
\bibitem [{\citenamefont {Cardoso}\ \emph {et~al.}(2022)\citenamefont
  {Cardoso}, \citenamefont {Destounis}, \citenamefont {Duque}, \citenamefont
  {Macedo},\ and\ \citenamefont {Maselli}}]{Cardoso:2021wlq}%
  \BibitemOpen
  \bibfield  {author} {\bibinfo {author} {\bibfnamefont {V.}~\bibnamefont
  {Cardoso}}, \bibinfo {author} {\bibfnamefont {K.}~\bibnamefont {Destounis}},
  \bibinfo {author} {\bibfnamefont {F.}~\bibnamefont {Duque}}, \bibinfo
  {author} {\bibfnamefont {R.~P.}\ \bibnamefont {Macedo}},\ and\ \bibinfo
  {author} {\bibfnamefont {A.}~\bibnamefont {Maselli}},\ }\bibfield  {title}
  {\bibinfo {title} {{Black holes in galaxies: environmental impact on
  gravitational-wave generation and propagation}},\ }\href
  {https://doi.org/10.1103/PhysRevD.105.L061501} {\bibfield  {journal}
  {\bibinfo  {journal} {Phys. Rev. D}\ }\textbf {\bibinfo {volume} {105}},\
  \bibinfo {pages} {L061501} (\bibinfo {year} {2022})}
  \BibitemShut
  {NoStop}%
\bibitem [{\citenamefont {Elizaga~Navacu\'es}, \citenamefont{Mena~Marug{\'a}n}\ and\ \citenamefont
  {M{\'\i}nguez-S{\'a}nchez}(2023)}]{BGA}%
  \BibitemOpen
  \bibfield  {author} {\bibinfo {author} {\bibfnamefont {B.}\ \bibnamefont
  {Elizaga~Navacu\'es}},\ \bibinfo {author} {\bibfnamefont {G.~A.}\ \bibnamefont
  {Mena~Marug{\'a}n}},\ and\ \bibinfo {author} {\bibfnamefont {A.}~\bibnamefont
  {M{\'\i}nguez-S{\'a}nchez}},\ }\bibfield  {title} {\bibinfo {title} {{Extended phase space quantization of a black hole interior model in loop quantum cosmology}},\
  }\href {https://doi.org/10.1103/PhysRevD.108.106001} {\bibfield  {journal}
  {\bibinfo  {journal} {Phys. Rev. D}\ }\textbf {\bibinfo {volume} {108}},\
  \bibinfo {pages} {106001} (\bibinfo {year} {2023})}
  \BibitemShut
  {NoStop}%
\end{thebibliography}
\end{document}